\documentclass[12pt]{iopart}

\usepackage{graphicx}
\usepackage{cite}
\usepackage{color}
\usepackage{multirow}
\usepackage[section]{placeins}

\begin{document}

\setlength{\parindent}{0pt}

\title[Ion energy and angular distributions in capacitive oxygen RF discharges]{Ion energy and angular distributions in low-pressure capacitive oxygen RF discharges driven by tailored voltage waveforms}

\author{Zolt\'an Donk\'o$^{1,2}$, Aranka Derzsi$^{1,3}$, M\'at\'e Vass$^1$, Julian Schulze$^{3,4}$ Edmund Schuengel$^5$, Satoshi Hamaguchi$^2$}

\address{$^1$Institute for Solid State Physics and Optics, Wigner Research Centre for Physics, Hungarian Academy of Sciences, 1121 Budapest, Konkoly Thege Mikl\'os str. 29-33, Hungary\\
$^2$Center for Atomic and Molecular Technologies, Graduate School of Engineering, Osaka University,
2-1 Yamadaoka, Suita, Osaka 565-0871, Japan\\
$^3$Department of Physics, West Virginia University, Morgantown, USA \\
$^4$Institute for Electrical Engineering, Ruhr-University-Bochum, Bochum, Germany \\
$^5$Evatec AG, 9477 Truebbach, Switzerland\\
}
\ead{donko.zoltan@wigner.mta.hu}

\begin{abstract}
We investigate the energy and angular distributions of the ions reaching the electrodes in low-pressure, capacitively coupled oxygen radio-frequency discharges. These distributions, as well as the possibilities of the independent control of the ion flux and the ion energy are analysed for different types of excitation: single- and classical dual-frequency, as well as valleys- and sawtooth-type waveforms. The studies are based on kinetic, particle-based simulations that reveal the physics of these discharges in great details. The conditions cover weakly collisional to highly collisional domains of ion transport via the electrode sheaths. Analytical models are also applied to understand the features of the energy and angular distribution functions.   
\end{abstract}

\submitto{\PSST}
\maketitle

\section{Introduction}

The active species created in discharge plasmas provide the basis of various surface modification techniques, such as etching, deposition, surface microstructuring and functionalisation \cite{G1,G2,G3,G4}. The species utilised in these applications are usually ions and radicals, which participate in different physical and chemical interactions with the surface layers. Controlling the  flux, the energy and the angle of incidence of these active species at the surfaces is of primary importance. In order to be able to regulate the processes taking place at the plasma-surface interface the connection between the operating conditions of the plasma and the resulting ion properties, such as the ion flux, the mean ion energy, the Ion Flux-Energy Distribution Function (IFEDF), and the Ion Angular Distribution Function (IADF), has to be understood. Both the formation and the characteristics of the IFEDF and the IADF \cite{Wild,Kawamura,Eddi15,Chen18,WangS,Hamaguchi,Kushner,Kratzer,Woodworth,Rakhimova,Huang,NEW2} received significant attention during the past decades. While the IFEDF is important in all applications, the IADF is highly relevant, e.g. in high aspect ratio etching \cite{Wu,Donnelly}. 

Capacitively Coupled Plasmas (CCPs) represent one of the most important sources used in the applications mentioned above. In these systems the ions flying towards the electrodes have to traverse the sheaths (over which a significant voltage drop is present) and the ion energy and angular distributions are established within these regions \cite{Wild,Kawamura,Eddi15,Chen18}, the collisionality of which is an important factor in forming these distributions. Low pressures and narrow sheath widths result in collisionless or nearly collisionless transfer and a narrow angular distribution. At the other extreme, when the sheaths are much longer than the free path of the ions, the transport is highly collisional. This results in low-energy ions arriving with a broad angular distribution at the surfaces. At low pressures, the other important parameter is the ratio of the ion transit time to the period of the applied radio-frequency (RF) waveform. Under conditions when this ratio is small, the ions acquire an energy that corresponds closely to the instantaneous sheath voltage, while at high values of this ratio, the ions fly through the sheath during several RF periods, and their energy is determined by the time-averaged sheath voltage. 

By changing the pressure, the driving frequency, and the voltage amplitude, a variety of IFEDFs can be realised  \cite{DonkoEPS}. Nonetheless, various ways for an additional control of the ion properties have been searched for during the past decades. Solutions were found in the extra degrees of freedom provided by the driving voltage waveform, which can be more complex than the "default" single harmonic signal.

An independent control of the ion properties (the mean ion energy and the ion flux) was first made possible by introducing {\it Dual-Frequency (DF) excitation} to drive capacitively coupled plasmas in 1992 \cite{Goto}. When significantly different driving frequencies are used, the plasma production and charged particle densities are primarily controlled by the amplitude of the high-frequency voltage, while the transport of the ions across the sheaths is primarily determined by the low-frequency voltage amplitude. Properties of plasma sources, operated in various gases and under different conditions, driven by DF waveforms have thoroughly been studied \cite{DF1,DF2,DF3,DF4}. These investigations have also revealed that the independent control of ion flux and energy is limited by "frequency coupling" effects \cite{coupling1,coupling2} and secondary electron emission from the electrodes \cite{secondaries1,secondaries2}. 

A further major step in the control of ion properties has been the discovery of the {\it Electrical Asymmetry Effect} (EAE) \cite{EAE,Heil2} in 2008, by using a base frequency and its second harmonic to excite the plasma. Such a driving voltage was shown to lead to the 
development of a DC self-bias even in geometrically symmetrical systems. This self-bias can be controlled by the phase between the two harmonics and has a direct effect on the energy of the ions at the electrodes, while the ion flux remains approximately constant \cite{DZ-EAE}. The performance of the EAE in various electropositive and electronegative gases has been investigated in details, including electron power absorption mechanisms and transitions  between them, the effects of the secondary electron emission and of the driving frequencies, etc. \cite{EAE-neg,EAE-freq,EAE-freq2,ZYR,EAE-general}. 

The possibility of using a higher number of harmonics, leading to {\it peaks- and valleys-type waveforms}, and the optimization of the harmonic voltage amplitudes have also been explored \cite{Eddi15,optim1,optim2,Wendt1,Wendt2}. These types of waveforms belong to the set of {\it Tailored Voltage Waveforms} (TVW, discussed in details in a comprehensive review \cite{Trevor}), which also cover {\it sawtooth-type waveforms} (introduced in 2014) \cite{sawtooth,sawtooth2}. Through the past years TVWs were utilised as well in practical applications, e.g. in silicon thin film deposition \cite{EAE-depos,EAE-depos2} and etching \cite{J_1,J_2}.

In this work we investigate oxygen plasmas. The choice of this gas is motivated by the complex physics of oxygen plasmas, as well as by the practical importance of this gas: oxygen plasmas have been used, e.g., in high-tech applications based on etching by reactive plasma species \cite{Coburn,Childres}, in patterning of highly oriented pyrolytic graphite \cite{Lu}, resist stripping for multilayer lithography \cite{Hartney}, in the modification of various coatings and films \cite{Cvelbar,Nakamura,Chou}, in the creation of micropatterns of chemisorbed cell adhesion-repellent films \cite{Tourovskaia,H1,H2} and in the production of highly porous SnO$_2$ fibers \cite{Zhang}. 

Oxygen discharges have been the subject of numerous fundamental discharge studies. The EAE in oxygen CCPs was studied experimentally and via particle simulations by Schuengel {\it et al.}\cite{Eddi-O2} and Zhang {\it et al.} \cite{Zhang-O2}. Emission patterns this is the last candidate.  next esc will revert to uncompleted text. hat reveal information about the spatio-temporal distribution of the excitation rates caused by different species have been analysed by Dittmann {\it et al.}  \cite{Dittmann}. The surface recombination of the singlet delta oxygen metastable molecules has been addressed in \cite{Greb,Greb2,NEW}. A comparison between experiments and simulations for peaks- and valleys-type waveforms, focusing on the DC self-bias, the ion flux, the discharge power, and the ion flux-energy distributions has been presented in \cite{Derzsi2016}. These investigations have been extended by Phase Resolved Optical Emission Spectroscopy (PROES) studies. Derzsi {\it et al.} \cite{Derzsi2017} presented a detailed comparison between experimental and simulation data for excitation maps for peaks- and valleys-type, as well as sawtooth-type waveforms. Transitions between the electron power absorption modes associated with the sheath dynamics ("$\alpha$-mode") and with the bulk and ambipolar electric fields ("Drift-Ambipolar" or "DA"-mode \cite{DAmode}) have been identified as a function of the operating conditions \cite{Derzsi2017,Gud2017}. A (limited) sensitivity analysis of some of the model parameters has been presented by Donk\'o {\it et al} \cite{DonkoEPS2018}, while the effect of the driving frequency has recently been analysed via Particle-in-Cell simulations incorporating Monte Carlo collisions (PIC/MCC) by Gudmundsson {\it et al.} \cite{G2018}. The effect of creation of negative ions at the electrode surfaces has been addressed in \cite{Matt}. These (and several other) studies have uncovered much of the effects taking place in low-pressure oxygen CCPs, however, the analysis of the formation of the ion angular and energy distribution functions in discharges driven by different voltage waveforms is not yet understood and warrants further investigations. 

Therefore, in this paper, we investigate the ion properties in low-pressure oxygen CCPs driven by various waveform types: single-frequency, classical dual-frequency, valleys and sawtooth. In particular, we focus on the flux-energy distributions, the angular distributions, as well as the joint energy and angular distributions of the impinging positive ions at the electrodes, as a function of (i) the driving voltage waveform, (ii) the gas pressure, and (iii) the ion-induced secondary electron emission coefficient (SEEC, $\gamma$). We present, as well, a simple model for the formation of the joint energy and angular distribution at conditions when the sheaths are weakly collisional and use a more elaborated model \cite{Eddi} to derive the IFEDF for any conditions. We also discuss general characteristics of oxygen plasmas such as electron power absorption modes, electronegativity, and behaviour of the DC self-bias under the conditions of excitation by TVWs. The discharge model (including the set of elementary processes and the definition of the driving voltage waveforms) is presented in section \ref{sec:model}. The results of our investigations are presented in section 3, while section 4 gives a brief summary of our studies.

\section{Discharge model, computational implementation, and discharge conditions}

\label{sec:model} 

Below, we present the main features of the discharge model and its computational implementation. Subsequently, we specify the discharge conditions, including the various driving voltage waveforms that are used in our studies of oxygen CCPs.

\subsection{Model of oxygen CCPs and computational implementation}

Our model of the oxygen CCPs is largely based on the "xpdp1" set of elementary processes \cite{Vahedi} and its recent revision \cite{Gudmundsson}, and is the same as described in our previous works \cite{Derzsi2016,Derzsi2017,DonkoEPS2018}. Therefore, only a brief summary of the features of the model is given below and the reader is referred to \cite{Derzsi2016} for details. 

The charged species considered in the model are O$_2^+$ and O$^-$ ions and electrons. The set of elementary collision processes between the electrons and O$_2$ neutral molecules includes elastic scattering, excitation to rotational, vibrational and electronic levels, ionisation, dissociative excitation, dissociative attachment, impact detachment, as well as dissociative recombination. For O$_2^+$ ions elastic collisions with O$_2$ are taken into account; we include the symmetric charge exchange process and an additional channel with isotropic scattering in the center-of-mass frame (the cross section for the isotropic channel is set to be 50\% of the charge exchange cross section as suggested in \cite{Gudmundsson}). For O$^-$ ions the model includes elastic scattering with O$_2$ neutrals, detachment in collisions with electrons and O$_2$ molecules, mutual neutralization with O$_2^+$ ions, as well as collisions with metastable singlet delta oxygen molecules, O$_{2}(a^{1}\Delta_{\rm g})$. This latter species is known to play an important role in oxygen CCPs \cite{Greb,Greb2,NEW}, especially in establishing the negative ion balance. In our model we assume that this species has a spatially uniform density that is computed from the balance between their creation rate in the gas phase by e$^-$+O$_2$ collisions and their loss rate at the electrode surfaces \cite{Derzsi2016}. 

Compared to the original xpdp1 set, we replace the elastic collision cross-section with the elastic momentum transfer cross-section of \cite{Biagi} and assume isotropic electron scattering, replace the xpdp1 ionisation cross-section with that recommended in \cite{Gudmundsson}, and adopt as well all the ion-molecule and ion-ion collision cross-sections from \cite{Gudmundsson}.

The model is implemented into a 1d3v Particle-in-Cell simulation code incorporating Monte Carlo treatment of collision processes (PIC/MCC) \cite{PIC1,PIC2,PIC3}. We assume plane and parallel electrodes with a gap of $L$ = 2.5 cm. One of the electrodes is driven by a voltage waveform (specified below), while the other electrode is at ground potential. Electrons are assumed to be elastically reflected from the electrodes with a probability of 0.2. The code allows including the emission of secondary electrons from the electrodes; we use SEEC values of $\gamma=0$ (disregard electron emission), $\gamma=0.06$ (a value that is characteristic for metal surfaces), and $\gamma=0.4$ (a value typical for dielectric surfaces). The gas temperature is fixed at $T_{\rm g}=350$ K. For the surface quenching probability of O$_{2}(a^{1}\Delta_{\rm g})$ singlet delta molecules we use the value of $\alpha = 6 \times 10^{-3}$, which resulted in our previous studies in a good overall agreement between the experimental and simulation data for the ion fluxes and the ion flux-energy distribution functions at the electrodes \cite{Derzsi2016,Derzsi2017,DonkoEPS2018}. 

In the cases when a DC self-bias voltage develops, its value is determined in an iterative manner to ensure equal losses of positive and negative charges at each electrode over one period of the fundamental driving frequency \cite{DZ-EAE}. This, self-consistently computed value is added to the excitation waveform applied at the powered electrode.

The computations are carried out using a spatial grid with $N_x$ = 100-1600 points and $N_t$ = 2000-85\,000 time steps within the fundamental RF period. These parameters have been set to fulfil the stability criteria of the computational scheme. We note that despite the relatively low number of different plasma species and elementary processes included in our model, our previous studies have concluded that this model is able to predict experimentally observable discharge characteristics with a reasonable accuracy, for a wide domain of conditions \cite{Derzsi2016,Derzsi2017,DonkoEPS2018}.  

Throughout this paper the distribution functions, viz. the ion flux energy distribution function, $F(\varepsilon)$, the ion angular distribution function, $F(\Theta)$, as well as the joint ion energy-angular distribution function, $F(\varepsilon,\Theta)$, represent the number of ions reaching the electrodes as a function of their energy ($\varepsilon$) and/or incidence angle ($\Theta$) \cite{Shihab}. This way, these distribution functions are "measured" in units of m$^{-2}$ s$^{-1}$ eV$^{-1}$, m$^{-2}$ s$^{-1}$ deg$^{-1}$, and m$^{-2}$ s$^{-1}$ eV$^{-1}$ deg$^{-1}$, respectively. It is noted that another definition of the angular distribution function that corresponds to the flux over solid angle is also commonly used; for a discussion see \cite{Shihab}.

\subsection{Excitation waveforms}

Simulations are conducted with various excitation waveforms, specified below. Table 1 lists these waveforms and  summarises the parameters used. Subsequently, the mathematical forms of the excitation voltage waveforms are given.

\begin{table}
\caption{\label{o2} Operating conditions of oxygen capacitive discharges studied in this paper. The peak-to-peak voltage is always $\phi_{\rm pp}$ = 400 V.}
\footnotesize
\begin{tabular}{@{}lllll}
\br
Driving waveform&Frequency / Frequencies & Pressure [Pa] & SEEC $\gamma$ & Waveform\\
\mr
Single harmonic & $f_1$ = 27.12 MHz & 5, 10, 20 & 0, 0.06, 0.4 & eq. (\ref{eq:exc1})\\
Classical dual frequency & $f_1$ = 27.12 MHz \& $f_2 = f_1 / 14$  & 5, 10, 20 & 0, 0.06, 0.4 & eq. (\ref{eq:exc2})\\
Peaks and valleys & $f_1$ = 15 MHz, $f_k = k f_1$  & 5, 10, 20 & 0, 0.4 & eq. (\ref{eq:exc3})\\
Sawtooth & $f_1$  = 15 MHz, $f_k = k f_1$  & 5, 10, 20 & 0, 0.4 & eq. (\ref{eq:exc4})\\
\br
\end{tabular}
\end{table}

The excitation waveforms for the different cases are:
\begin{itemize}
\item{single-frequency excitation: 
\begin{equation}
\phi(t) = \phi_1 \cos(2 \pi f_1 t),
\label{eq:exc1}
\end{equation}}
\item{classical dual-frequency excitation: 
\begin{equation}
\phi(t) = \phi_1 \cos(2 \pi f_1 t) + \phi_2 \cos(2 \pi f_2 t),
\label{eq:exc2}
\end{equation}
}
\item{peaks- and valleys-type waveforms: 
\begin{equation}
\phi(t) = \sum_{k=1}^N \phi_k \cos(2 \pi k f_1 t + \theta_k),
\label{eq:exc3}
\end{equation}
where $\theta_k$ are the phase angles and $N$ is the number of harmonics (for which we use a maximum number of 4). The $\phi_k $ amplitudes of the individual harmonics are set according to
\begin{equation}
\phi_k = \frac{2(N-k+1)}{(N+1)^2} \phi_{\rm pp},
\label{eq:exc3a}
\end{equation}
where $\phi_{\rm pp}$ is the peak-to-peak voltage. The peaks-type voltage waveforms can be realised by setting all $\theta_k$ angles to zero, while for the valleys-type waveforms the phase angles of all even harmonics have to be set to $\pi$. We note that by "reversing" the waveform (peaks $\leftrightarrow$ valleys), in the geometrically symmetrical configuration considered here the plasma is mirrored with respect to the centre of the electrode gap. (Therefore, it is sufficient to study discharges with one of the waveform types; we chose the valleys-type waveform in the present study.)}
\item{sawtooth-type waveforms: 
\begin{equation}
\phi(t) = \pm \phi_{\rm a} {\sum_{k=1}^N \frac{1}{k}\sin(2 \pi k f_1 t)},
\label{eq:exc4}
\end{equation}
where the minus and plus signs define, respectively, the sawtooth-up and sawtooth-down waveforms. The value of the $\phi_{\rm a}$ prefactor is set (for each value of the number of harmonics, $N$) in a way that the waveform has the specified peak-to-peak voltage, $\phi_{\rm pp}$.}
\end{itemize}

\begin{figure}[ht!]
\begin{center}
\includegraphics[width=\textwidth]{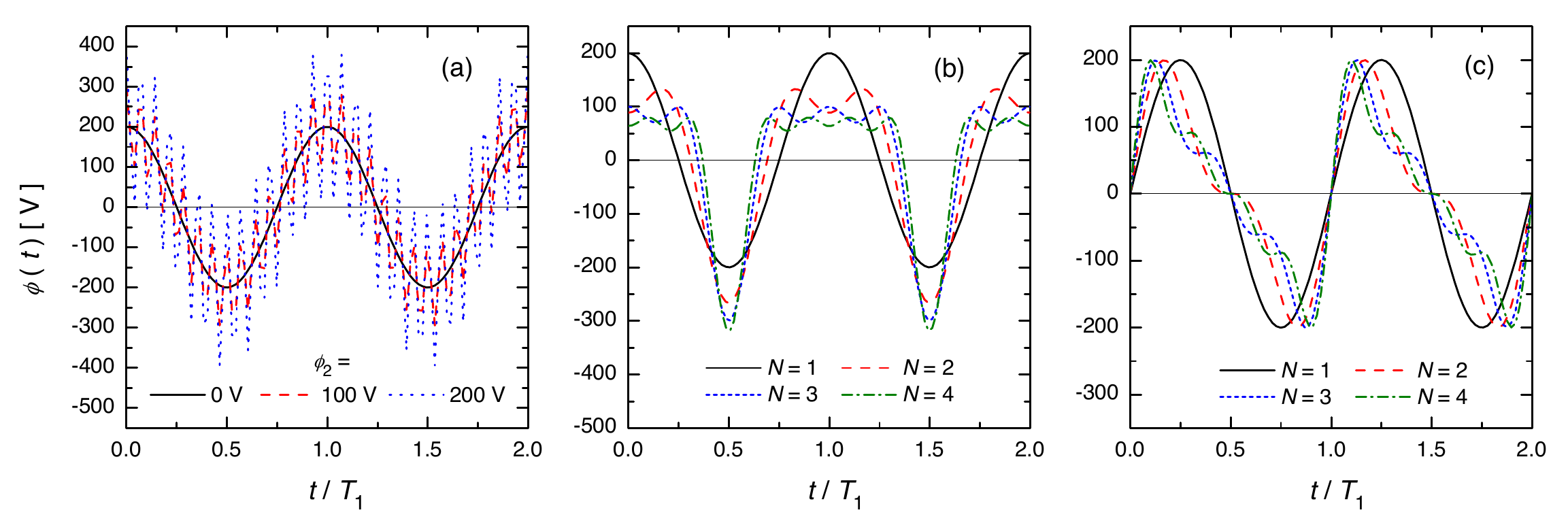}
\caption{Driving voltage waveforms covered in this work: (a) single- ($\phi_2$ = 0 V) and classical dual-frequency ($\phi_2 >$ 0 V) waveforms with $\phi_1$ = 200 V, (b) valleys-type and (c) sawtooth-type waveforms with different number of harmonics ($N$), for $\phi_{\rm pp}$~=~400~V. $T_1$ is the period of the fundamental frequency $f_1$.}
\label{fig:waves}
\end{center}
\end{figure}

Examples of driving voltage waveforms specified above are displayed in figure~\ref{fig:waves}.

\section{Results}

\subsection{Single-frequency excitation -- the general behaviour of the discharge}

\begin{figure}[ht!]
\begin{center}
\includegraphics[width=0.85\textwidth]{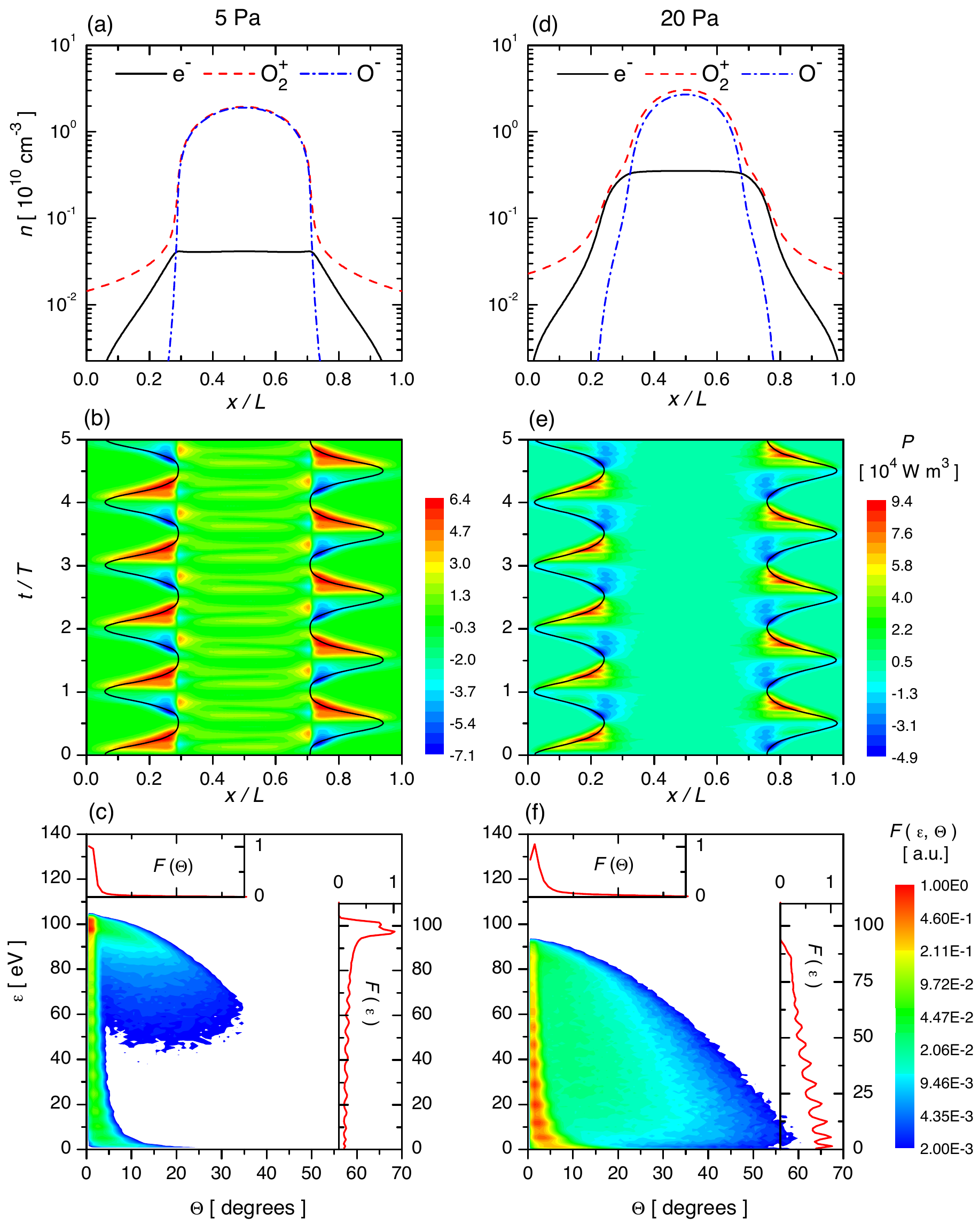}
\caption{Characteristics of single-frequency ($f_1=$~27.12 MHz) oxygen discharges at 5 Pa (left) and 20 Pa (right) pressures. Panels (a, d) show the time-averaged density distributions of the charged species, while (b, e) display the spatio-temporal electron power absorption rate for five RF periods ($T$). In panels (c, f) the energy and angular distribution of the O$_2^+$ ions, $F(\varepsilon,\Theta)$, is given (in arbitrary units) at the electrodes. The insets illustrate the energy and angular distributions (respective integrals of $F(\varepsilon,\Theta)$ according to incidence angle and energy). All distributions have been normalised to a maximum value of 1, for easier comparison. $\phi_{\rm pp}$ = 400 V, $\gamma=0$. The black lines in panels (b) and (e) mark the sheath edges.}
\label{fig:singlefreq1}
\end{center}
\end{figure}

First, we address the discharge behaviour under the simplest form of excitation, driving the plasma with a single-frequency harmonic voltage source given by eq.\,(\ref{eq:exc1}), with $f_1=$~27.12 MHz. Figure \ref{fig:singlefreq1} presents the simulation results for pressure values of 5~Pa and 20 Pa, for $\phi_{\rm pp}$ = 400 V peak-to-peak driving voltage, $L$ = 2.5 cm electrode gap, and an ion-induced secondary electron emission coefficient $\gamma=0$. Panels (a) and (d) show the time-averaged density distributions of the charged species. The discharge is highly electronegative at both pressures, the electronegativity decreases with increasing pressure, but the peak negative ion density remains approximately an order of magnitude higher than the peak electron density even at 20 Pa. (We define the electronegativity ($\beta$) as the ratio of the spatially averaged densities of negative ions and electrons.) A similar dependence of the electronegativity of the plasma on the oxygen pressures was also observed in \cite{Eddi-O2} and \cite{LVL}. Panels (b) and (e) of figure \ref{fig:singlefreq1} depict the electron power absorption rate, space- and time-resolved (for five radiofrequency periods). The power absorption peaks in both cases in the vicinity of the expanding sheath edges. The position of the sheath edges is found by the criterium defined by Brinkmann \cite{Brinkmann}. Taking as an example the sheath adjacent to the powered electrode situated at $x=0$, the position of the sheath edge, $s$, is found from
\begin{equation}
\int_0^s n_{\rm e}(x) {\rm d} x = \int_s^{x^\ast} [n_{\rm i}(x)- n_{\rm e}(x)] {\rm d} x,
\end{equation}
where $n_{\rm i}$ and $n_{\rm e}$ are, respectively, the total positive and negative charged particle densities, and $x^\ast$ is a position towards the centre of the discharge where quasineutrality holds.

Power absorption in the bulk plasma domain and at the edges of the collapsing sheaths, which is a sign of the Drift-Ambipolar power absorption mode \cite{DAmode}, is revealed at the lower pressure of 5 Pa only. In this latter case the mean energy of the electrons in the centre of the plasma is 6.3 eV, whereas in the 20 Pa case it is only 2.1 eV. The high value of the mean electron energy at the lower pressure is required because of the need to conduct the current by the electrons having a depleted density. At higher pressure, where the electron density in the bulk is about an order of magnitude higher with respect to the 5 Pa case (the electronegativity decreases), such a high electron energy is not required and the mean electron energy decays significantly.

We note that oxygen plasmas are highly electronegative at low pressures, while discharges operated in, e.g., CF$_4$, are highly electronegative at high pressures and much less electronegative at low pressures. In O$_2$, detachment due to collisions of negative ions with metastable molecules causes the decrease of the electronegativity as a function of pressure (as confirmed by our simulation results, not discussed here in details). Such different behaviour of electronegative plasmas can lead to different modes of discharge operation with drastic consequences for process control.

The $F(\varepsilon,\Theta)$ flux energy and angular distribution of the O$_2^+$ ions upon their impingement at the electrodes is displayed in figures~\ref{fig:singlefreq1}(c) and (f), for the two pressures studied. Insets of these panels show the flux energy distribution $F(\varepsilon)$ (integral of $F(\varepsilon,\Theta)$ over all angles) and the flux angular distribution $F(\Theta)$ (integral of $F(\varepsilon,\Theta)$ over all energies). At low pressure (5 Pa) most of the ions arrive at angles smaller than a few degrees and the energy distribution peaks at $\varepsilon \approx$ 100 eV, that corresponds to the time average of the sheath voltage. This indicates a weakly collisional sheath, which the ions traverse, however, over several RF cycles. At the higher pressure of 20 Pa the $F(\varepsilon,\Theta)$ distribution reveals several discrete spots that originate from the charge exchange collisions, which, in the presence of a periodically varying sheath electric field, synchronise the motion of groups of ions. (Ions that have undergone a charge exchange collision at times of low electric field "gather" until the field grows again and moves these ions -- for details see \cite{Wild,Eddi}.) In general, the angular distribution of the incoming ions broadens and the energy distribution, due to the higher collisionality of the electrode sheaths, becomes a decreasing function of the ion energy (except of its peaks originating from charge exchange collisions).  

\begin{figure}[ht!]
\begin{center}
\includegraphics[width=0.5\textwidth]{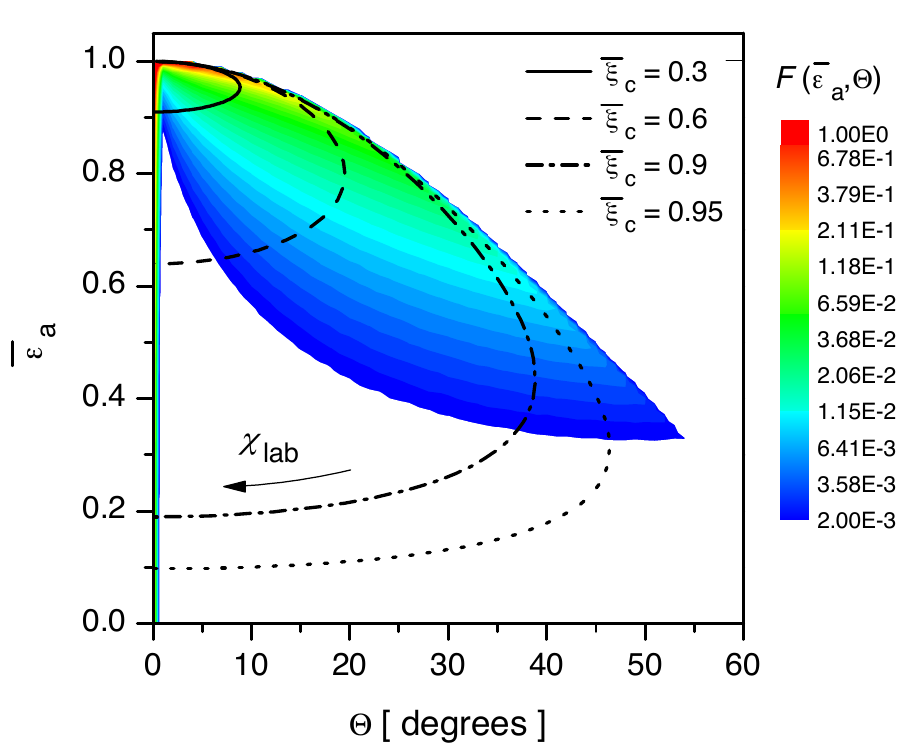}
\caption{The colour map shows the $F(\overline{\varepsilon}_{\rm a},\Theta)$ distribution of the O$_2^+$ ions arriving at the electrodes, as obtained from the model. The distribution has been normalised to a maximum value of 1. The lines show the loci of the possible $(\overline{\varepsilon}_{\rm a},\Theta)$ pairs after an ion collision at different normalised positions, $\overline{\xi}_{\rm c}$, (see text) within the sheath. The arrow shows the direction of an increasing $\chi_{\rm lab}$ from 0$^\circ$ to 90$^\circ$ along these lines.}
\label{fig:mushroom}
\end{center}
\end{figure}

The "mushroom" shape of the $F(\varepsilon,\Theta)$ distribution at low pressure (e.g. 5 Pa, figure~\ref{fig:singlefreq1}(c)) can be understood based on an analytical model of the motion of O$_2^+$ ions. In this model we assume that each ion undergoes {\it exactly one collision} within the sheath, at a {\it random position}. The energy of the ions entering the sheath is assumed to be zero. As the ion transit time through the sheath is significantly longer than the RF period, ions can be considered to sense the time-averaged electric field. The latter is taken to be linear with respect to the position within the sheath. We consider the grounded sheath and fix the zero of our coordinate system ($\xi=0$) at the position of the maximum sheath width. At this point we also set the electric field and the potential to zero, $E = 0$ V/m, $\Phi=0$ V. Under these conditions the electric field and the potential within the sheath vary as:
\begin{eqnarray}
E(\xi) = -\frac{2 \phi_{\rm s}}{s^2} \, \xi= -\frac{2 \phi_{\rm s}}{s} \,\overline{\xi} \\
\Phi(\xi) = \Phi_{\rm s} \biggl( \frac{\xi}{s} \biggr)^2 = \Phi_{\rm s} \, \overline{\xi}^2,
\end{eqnarray}
where $\Phi_{\rm s}$ is the time-averaged sheath voltage, $s$ is the maximum width of the sheath, and $\overline{\xi}= \xi / s$ is the normalised spatial coordinate within the sheath. If the ion undergoes a collision at $\overline{\xi}_{\rm c}$, the pre- and post-collision energies are, respectively,
\begin{eqnarray}
\varepsilon_{\rm c} = Q \, \Phi_{\rm s} \, \overline{\xi}^2 \\
\varepsilon_{\rm s} = \varepsilon_{\rm c} - \Delta \varepsilon = \varepsilon_{\rm c} \cos^2 \chi_{\rm lab} = Q \, \Phi_{\rm s} \,\overline{\xi}^2  \cos^2 \chi_{\rm lab},
\end{eqnarray}
as the relative change of the energy is $\Delta \varepsilon / \varepsilon_{\rm c}  = (1- \cos \chi_{\rm com}) /2$ and $\chi_{\rm lab} = \chi_{\rm com}/2$, because the masses of the collision partners are the same. Here $\chi_{\rm com}$ and $\chi_{\rm lab}$ are the scattering angles in the center-of-mass (COM) and laboratory (LAB) frames, respectively, and $Q$ is the charge of the ion. The energy of the ion upon arrival at the electrode is:
\begin{equation}
\varepsilon_{\rm a} = \varepsilon_{\rm s} + Q \, \bigl[  \Phi_{\rm s} -  \Phi(\overline{\xi}_{\rm c}) \bigr] = Q \, \Phi_{\rm s} \bigl[ 1 - \overline{\xi}_{\rm c}^2 (1 - \cos^2 \chi_{\rm lab}) \bigr].
\end{equation}
Introducing $\overline{\varepsilon}_{\rm a}$ as the arrival energy normalised by its maximum possible value $\varepsilon_{{\rm a,max}} = Q \, \Phi_{\rm s}$, we obtain 
\begin{equation}
\overline{\varepsilon}_{\rm a} = 1 -\overline{\xi}_{\rm c}^2 (1 - \cos^2 \chi_{\rm lab}).
\label{eq:model_energy}
\end{equation}
The angle of incidence can be computed from the perpendicular component of the post-collision velocity (that does not change along the post-collision trajectory), $v_{{\rm s},\perp}$, and the magnitude of the velocity upon arrival, $v_{\rm a}$, as
\begin{equation}
\sin \Theta = \frac{v_{{\rm s},\perp}}{v_{\rm a}} = \frac{\overline{\xi}_{\rm c} \sin (2 \chi_{\rm lab})}{2 \sqrt{\overline{\varepsilon_{\rm a}}}}.
\label{eq:model_angle}
\end{equation}

\begin{figure}[ht!]
\begin{center}
\includegraphics[width=0.45\textwidth]{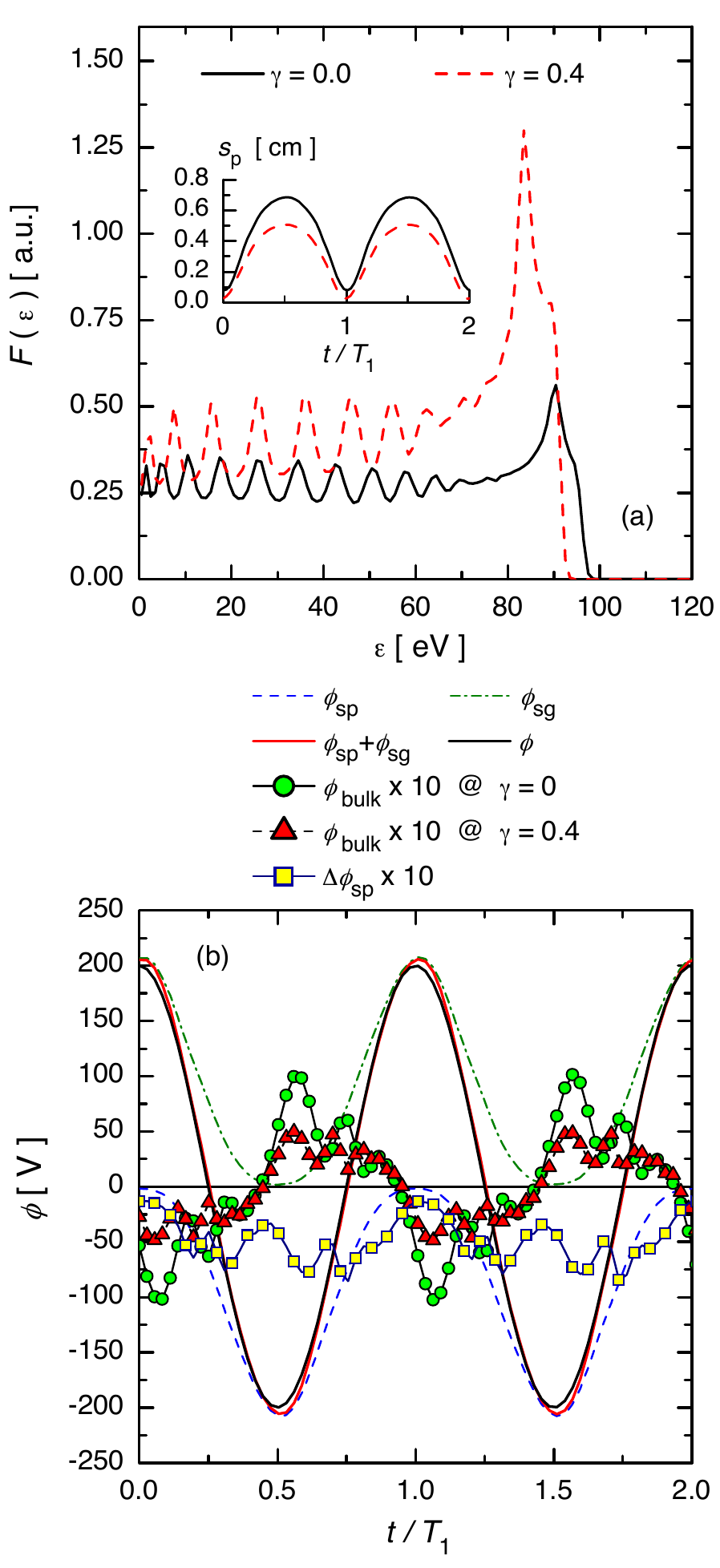}
\caption{(a) The effect of the SEEC $\gamma$ on the flux-energy distribution of O$_2^+$ ions, at $p$ = 10 Pa  ($f_1=$~27.12 MHz, $\phi_{\rm pp}$ = 400 V). The inset shows the time dependence of the width of the powered sheath, $s_{\rm p}$. (b) Time dependence of the sheath voltages ($\phi_{\rm sp}$, $\phi_{\rm sg}$) and their sum for $\gamma=0$, the applied voltage waveform ($\phi$), the bulk voltage drop ($\phi_{\rm bulk}$) for the $\gamma=0$ and $\gamma=0.4$ cases, and the difference between the voltage drops over the powered sheath between the two cases of $\gamma=0$ and $\gamma=0.4$ ($\Delta \phi_{\rm sp}$). Note that the $\phi_{\rm bulk}$ and  $\Delta \phi_{\rm sp}$ values are multiplied by a factor of 10. $T_1$ is the period of the driving frequency.}
\label{fig:single_idf_pot}
\end{center}
\end{figure}

Equations (\ref{eq:model_energy}) and (\ref{eq:model_angle}) give the normalised arrival energy and the incidence angle at the electrode for any position of collision, $\overline{\xi}_{\rm c}$, and laboratory scattering angle, $\chi_{\rm lab}$. The $F(\overline{\varepsilon}_{\rm a},\Theta)$ distribution can be simulated by generating a high number of random samples of these two variables. The result of such a simulation is shown in figure~\ref{fig:mushroom}. We note that in this simulation the collisions were executed by the same approach as in the PIC/MCC simulations: charge exchange (backward scattering in the COM system) was supposed to occur with a probability of 2/3 and isotropic scattering (in the COM system) was supposed to occur with a probability of 1/3. The results of the computation resemble closely the PIC/MCC simulation result shown in figure~\ref{fig:singlefreq1}, despite the very simple nature of the model. The formation of the main feature seen in the $F(\overline{\varepsilon}_{\rm a},\Theta)$ colour map of figure~\ref{fig:mushroom} can be understood by considering scattering events at specific $\overline{\xi}_{\rm c}$ positions. Figure~\ref{fig:mushroom} shows sets of $(\overline{\varepsilon},\Theta)$ pairs (forming lines) at fixed values of $\overline{\xi}_{\rm c}$. These lines were generated by scanning $\chi_{\rm lab}$ between 0$^\circ$ and 90$^\circ$. The $F(\overline{\varepsilon}_{\rm a},\Theta)$ distribution is nothing else but a superposition of an infinite number of such sets having infinitesimal widths. The model, on the other hand, cannot reproduce the low-energy part of the distribution obtained in the PIC/MCC simulation, which is a result of multiple collision (isotropic scattering or charge exchange) events.

The effect of the SEEC $\gamma$ on the flux-energy distribution of the O$_2^+$ ions is illustrated in figure~\ref{fig:single_idf_pot}(a) by PIC/MCC results. One can observe several differences between the two curves, corresponding to $\gamma$ = 0 and $\gamma=0.4$. First, the integral of $F(\varepsilon)$ increases as $\gamma$ is increased. This is attributed to an enhanced plasma density, that, in turn, results in shorter sheaths and lower collisionality, in the presence of appreciable secondary electron emission. (Note that while figure~\ref{fig:single_idf_pot}(a) presents the $F(\varepsilon)$ data in arbitrary units, the data are proportional to the real fluxes in the two cases). Second, $F(\varepsilon)$ exhibits fewer peaks at higher $\gamma$, which is the consequence of the fewer number of charge exchange collisions under such conditions, caused by the decrease of the sheath widths; see the time dependence of the width of the powered sheath, $s_{\rm p}$ in the inset of figure~\ref{fig:single_idf_pot}(a). 

The third difference between the $F(\varepsilon)$ curves computed at $\gamma$ = 0 and 0.4 is that the maximum ion energy decreases and the dominant peak of $F(\varepsilon)$ shifts towards lower energies by $\approx$~5 eV at $\gamma=0.4$. The origin of this shift is analysed in figure~\ref{fig:single_idf_pot}(b) that shows several voltage components: the voltage drops over the powered and the grounded sheaths ($\phi_{\rm sp}$ and $\phi_{\rm sg}$, respectively) and the sum of them for $\gamma=0$, the applied voltage waveform ($\phi$), as well as the bulk voltage drop ($\phi_{\rm bulk}$) for $\gamma=0$ and $\gamma=0.4$. (These voltages can be determined in a straightforward way when the sheath edge positions, $s_{\rm p}$ and $s_{\rm g}$ are known.) The figure also shows the difference of the voltage drops over the powered sheath in the two cases with $\gamma=0$ and $\gamma=0.4$, $\Delta \phi_{\rm sp} = \phi_{\rm sp,\gamma=0} -  \phi_{\rm sp,\gamma=0.4}$. Note that by definition $\phi_{\rm sp} < 0$ \cite{optim1}. 

The data reveal that $\Delta \phi_{\rm sp}(t)$ is negative at all times, i.e. the magnitude of the sheath voltage gets lower when $\gamma$ is increased. The temporal average of $| \Delta \phi_{\rm sp} |$ is $\approx$ 5 V, which explains the shift of the high-energy part of the $F(\varepsilon)$ distribution. Note that while we discussed the case of the powered sheath, the same arguments also hold for the grounded sheath. It is also to be mentioned that the sheath voltages do not add up to the applied voltage waveform, as a significant voltage drop over the bulk plasma $\phi_{\rm bulk} = \phi - (\phi_{\rm sp} +\phi_{\rm sg})$ forms, characteristic for electronegative plasmas. This voltage drop is actually, found to be responsible for the change of $\phi_{\rm sp}$, as the peak value of $| \phi_{\rm bulk} (t) |$ of $\approx$ 10 V at $\gamma=0$ decreases to $\approx$ 5 V at $\gamma=0.4$. This change is in turn, caused by the decreasing electronegativity with increasing $\gamma$: in the presence of an additional (surface) supply of electrons the conductivity of the bulk plasma increases and this requires a lower $\phi_{\rm bulk}$. As a secondary reason the decrease of the floating potential (by about 1.5 V) with increasing $\gamma$ can be identified that also leads to a decrease of the sheath voltage drop that is available to accelerate the positive ions. The effect of the SEEC on the "cutoff ion energy" is similar at other pressures, too. For both 5 Pa and 20 Pa we find lowering of the energy limit by a similar magnitude as was found and discussed for the case of 10 Pa.

\subsection {Classical dual-frequency excitation}

\label{sec:dual}

Next, we address the properties of oxygen CCPs under classical DF excitation as specified by eq.\,(\ref{eq:exc2}). Results will be presented for the high-frequency voltage amplitude $\phi_1$ set to 200 V and using a low-frequency voltage amplitude between $\phi_2$ = 0 V and 300~V. For some conditions the range of $\phi_2$ was limited by the adverse effect of $\phi_2$ on the charged particle densities, especially at low values of the SEEC.

\begin{figure}[ht!]
\begin{center}
\includegraphics[width=0.5\textwidth]{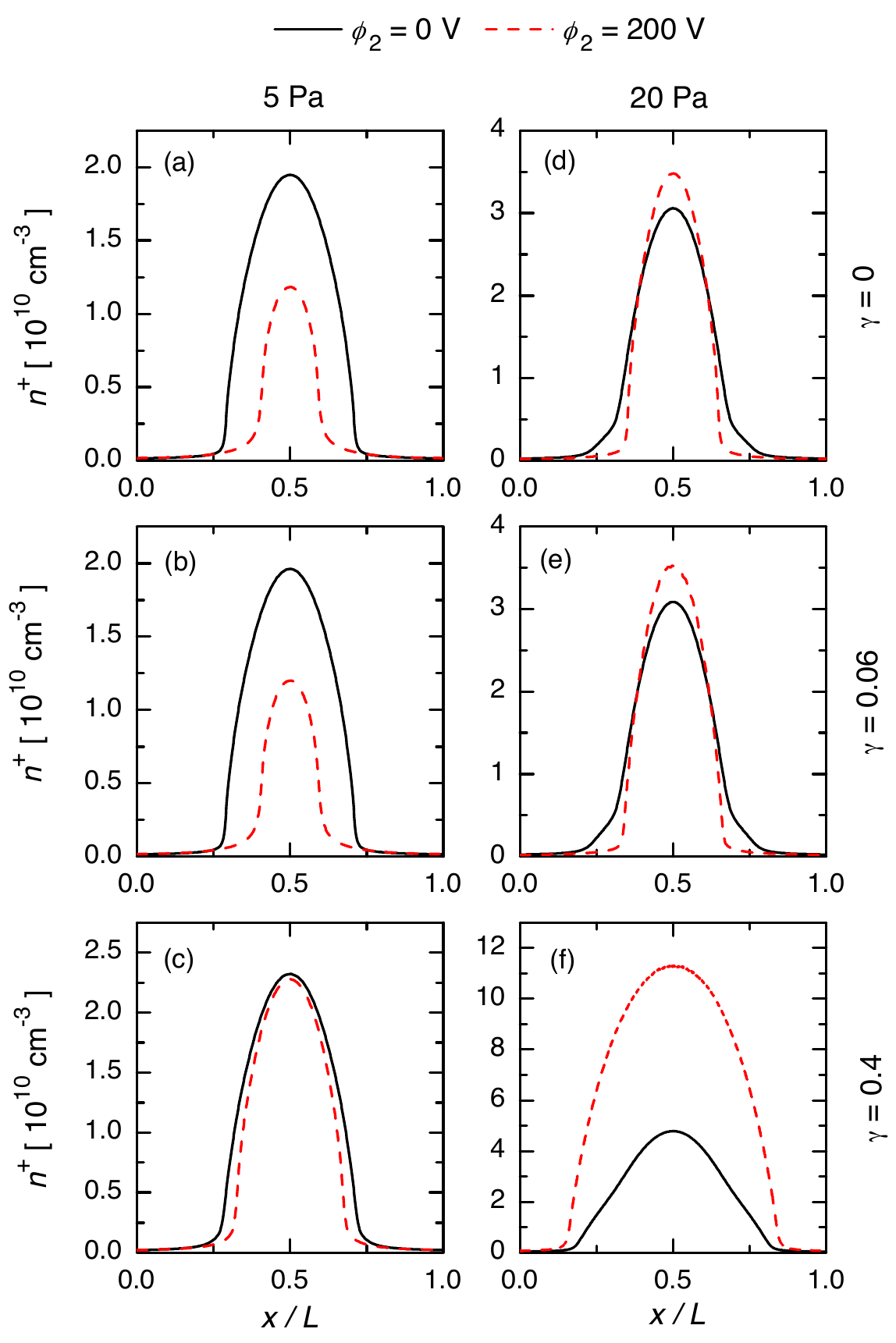}
\caption{The effect of the low-frequency voltage amplitude ($\phi_2$) on the density of O$_2^+$ ions in classical dual-frequency discharges at 5 Pa (left column) and 20 Pa (right column), at the $\gamma$ values specified. $\phi_1$ = 200 V.}
\label{fig:DF_densities}
\end{center}
\end{figure}

\begin{figure}[ht!]
\begin{center}
\includegraphics[width=0.5\textwidth]{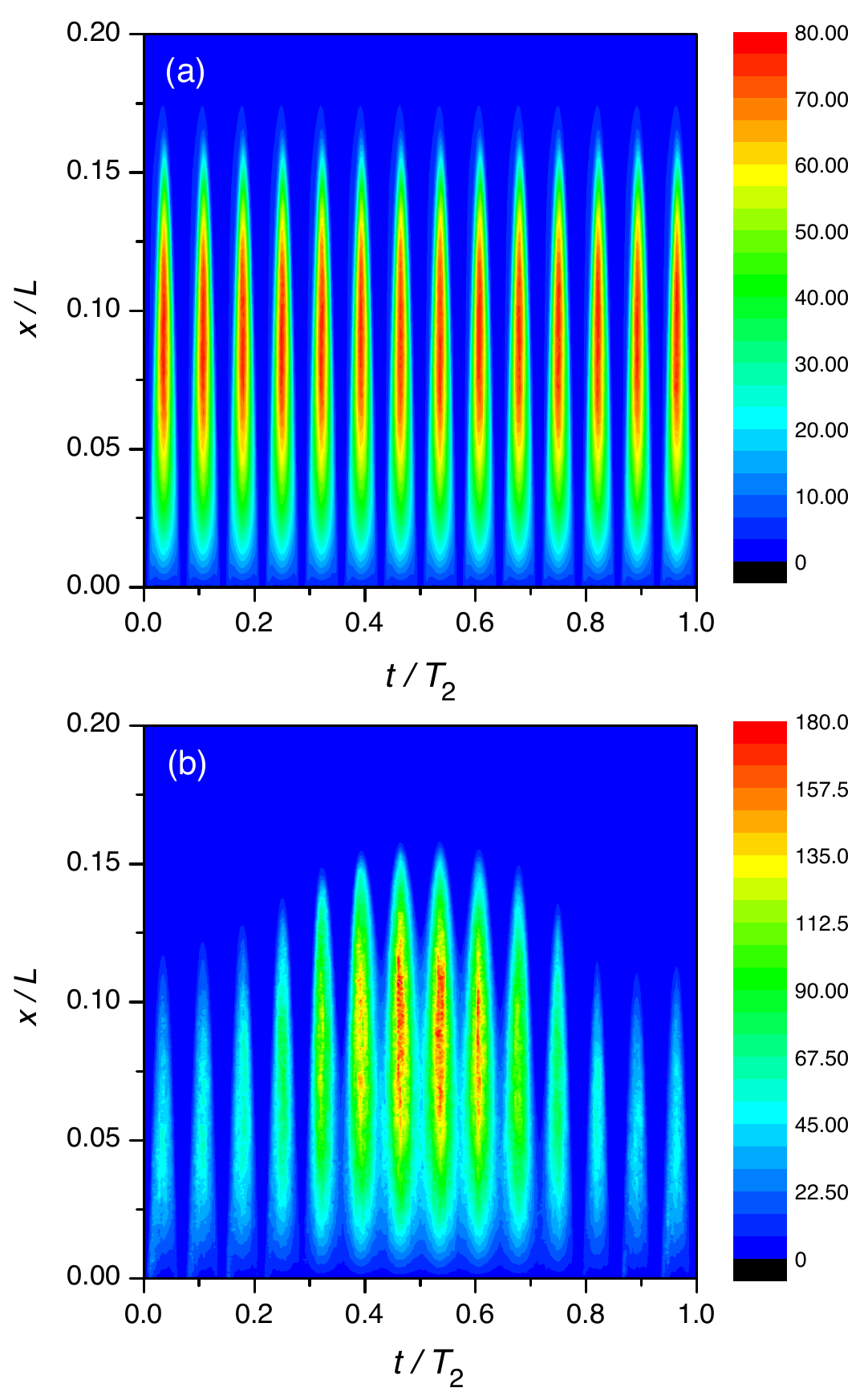}
\caption{The spatio-temporal variation of the mean electron energy in the vicinity of the powered electrode for $p$ = 20 Pa, $\phi_1$ = 200 V, and $\gamma=0.4$, at (a) $\phi_2$ = 0 V and (b) $\phi_2$ = 200 V. $T_2$ is the period of the low-frequency excitation.}
\label{fig:meane-xt}
\end{center}
\end{figure}

\begin{figure}[ht!]
\begin{center}
\includegraphics[width=0.4\textwidth]{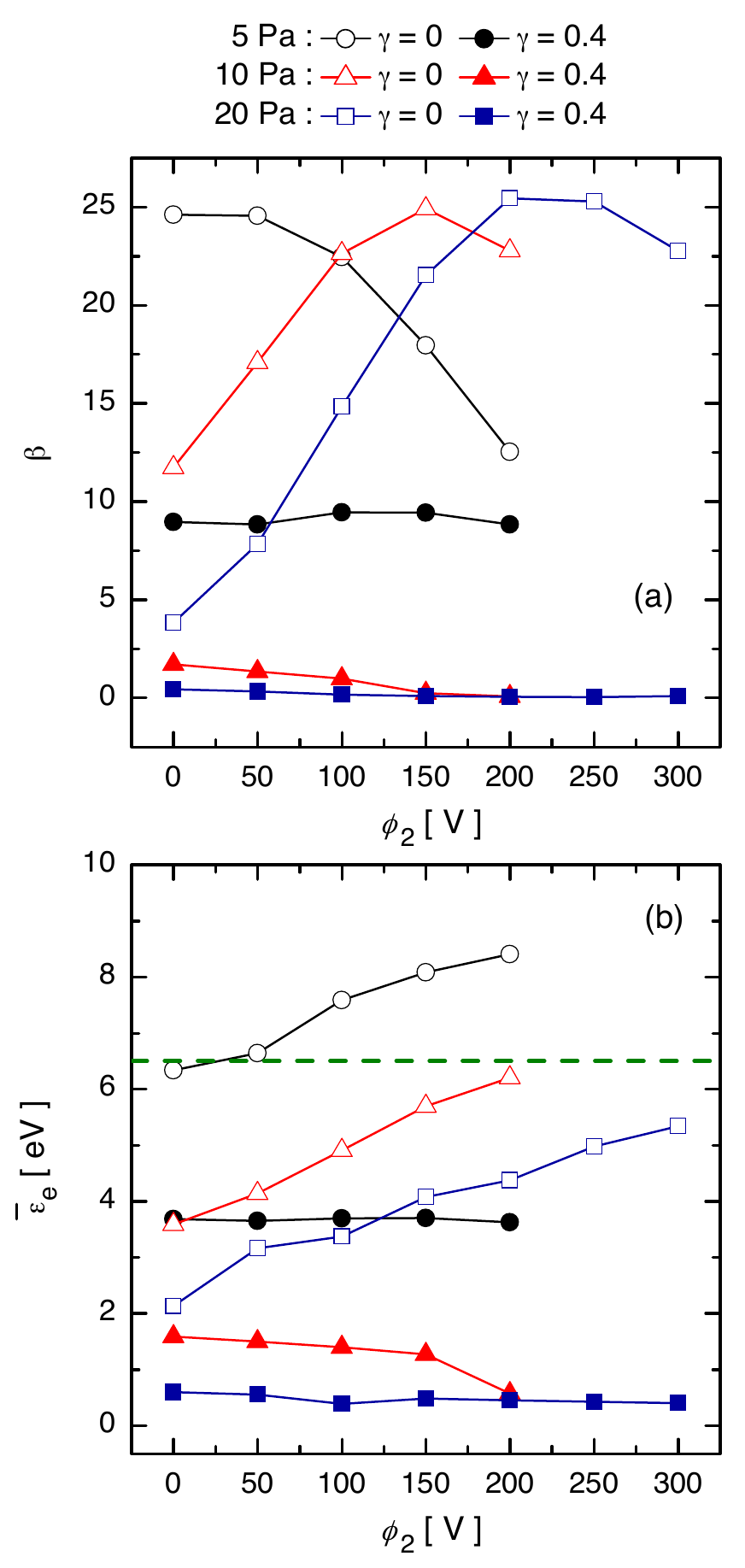}
\caption{The electronegativity, $\beta$, of the plasma (a) and the mean electron energy, $\overline{\varepsilon}_{\rm e}$, in the centre of the bulk (b), for classical dual-frequency excitation, as a function of $\phi_2$ for different pressures and SEEC values. $\phi_1$ = 200 V for all cases. The dashed green horizontal line in (b) indicates the energy (6.5 eV) where the attachment cross section peaks.}
\label{fig:DF_electronegativity}
\end{center}
\end{figure}

The spatial distribution of the density of O$_2^+$ ions is shown in figure~\ref{fig:DF_densities}, for different values of the gas pressure and the SEEC ($\gamma$). Each panel compares the distributions obtained at $\phi_2$ = 0 V and 200 V. At low pressure (5 Pa) the presence of the low-frequency component in the driving waveform results in a significant decrease of the ion density at zero / low $\gamma$ values (figures~\ref{fig:DF_densities}(a),(b)). This is a consequence of the widening of the sheath due to the higher voltage, which, on the other hand, does not contribute significantly to higher ionisation because of its low frequency, i.e. the frequency coupling mechanism \cite{coupling1,coupling2}.  At the highest value of the SEEC, $\gamma=0.4$, the enhanced sheath voltage contributes to the ionisation as it accelerates secondary electrons to high energies, which can then ionise as well. This effect almost exactly compensates the effect described above, leaving the ion density at the same value as in the case of $\phi_2$ = 0 V (see figure~\ref{fig:DF_densities}(c)). At 20 Pa pressure, as shown in  figures~\ref{fig:DF_densities}(d) and (e) for the lower $\gamma$ values, the sheath widths also slightly decrease and the ion density profiles become narrower and the peak ion density somewhat increases when $\phi_2$ = 200 V is applied. 

Comparing the cases shown in figures \ref{fig:DF_densities}(a) and (d) we observe an opposite effect of $\phi_2$ on the ion density at 5 Pa and 20 Pa. The application of the low-frequency voltage increases the sheath width in both cases, and this results in the reduction of the space available for the electrons to dissipate their energy acquired at sheath expansion. At 20 Pa the electrons dissipate all their energy in the bulk, whereas this is not the case at 5 Pa, where the electron mean free path is significantly longer.

At high $\gamma$ (figure~\ref{fig:DF_densities}(f)) a significant enhancement of the ion density is observed due to the effect of an efficient multiplication and an enhanced mean energy of the secondary electrons. The behaviour of the latter quantity is shown in figure \ref{fig:meane-xt}, for 20 Pa pressure, $\gamma$ = 0.4, for $\phi_2$ = 0 V (panel (a)) and $\phi_2$ = 200 V (panel (b)). These results indicate for  $\phi_2$ = 200 V a significant ($\sim$ factor of two) increase of the maximum value of the mean electron energy in space and time, with respect to the $\phi_2$ = 0 V case, when both $\phi_1$ and $\phi_2$ take their maximum values simultaneously in time. 

The specific driving voltage waveform has a significant effect as well on the electronegativity ($\beta$) of the plasma. The results are shown in figure~\ref{fig:DF_electronegativity}(a), for different pressures and SEEC values. In the case of $\gamma=0$, at low pressure (5 Pa) there is a remarkable decrease of the electronegativity with increasing $\phi_2$, whereas, for the higher pressures (10 Pa and 20 Pa) the opposite trend is observed: the increasing $\phi_2$ results in higher electronegativity. This behaviour of $\beta$ is discussed together with that of the mean electron energy in the centre of the plasma, $\overline{\varepsilon}_{\rm e}$, shown in figure~\ref{fig:DF_electronegativity}(b). 
At $\phi_2$ = 0 V, there is a strong correlation between a higher electronegativity and a higher mean electron energy, due to the reasons explained in the case of single-frequency discharges. When $\phi_2$ is increased, the same correlation is maintained at $p$ = 10 Pa, both the electronegativity and the mean electron energy increase with $\phi_2$ at zero $\gamma$ and decrease at high $\gamma$. At 20~Pa pressure the increase of the mean energy is less significant than the increase of the electronegativity, and at 5 Pa an opposite trend is observed at $\gamma=0$. The dependence of these observed trends on the pressure can be understood by noting that the electron attachment cross sections peaks at $\cong$ 6.5 eV, indicated by a horizontal line in the plot showing the mean electron energy. (This cross section has a sharp maximum around this value and has another broad peak above 20\,eV, with a smaller magnitude \cite{Vahedi}. This second feature of the cross section is not expected to influence our discussion.)

The application of the low-frequency voltage results in an increase of $\overline{\varepsilon}_{\rm e}$ at $\gamma=0$ (see figure~\ref{fig:DF_electronegativity}(b)). When the mean energy is lower than the energy corresponding to the peak of the electron attachment cross section, the attachment rate and, thus, the electronegativity increases with $\phi_2$. This is the scenario at 10 Pa and 20 Pa pressures. Here, the increased attachment rate results in an increased electric field within the bulk plasma due to the depleted electron density, that in turn, increases $\overline{\varepsilon}_{\rm e}$, i.e. there is a positive feedback mechanism leading to the establishment of the observed operating conditions. At 5 Pa, however, the mean energy $\overline{\varepsilon}_{\rm e}$ is near 6.5 eV at $\phi_2 =0$ V and its increase at $\phi_2 > 0$ V decreases the efficiency of the attachment and results in a decrease of the electronegativity, as seen in figure~\ref{fig:DF_electronegativity}(a). 

The non-monotonic behaviour of the electronegativity at $p$ = 10 Pa and 20 Pa with increasing $\phi_2$ at $\gamma=0$ is a result of two competing mechanisms. (i) As the mean electron energy approaches the value corresponding to the optimum of negative ion formation, the negative ion density increases in the bulk plasma. (ii) At the same time, also caused by the increase of the low-frequency voltage, the width of the bulk decreases, this way limiting the domain where negative ions can accumulate. As the electronegativity is defined "globally", i.e. as the ratio of the {\it spatially averaged} negative ion density and the {\it spatially averaged} electron density, the interplay of these two effects causes a transition, in our case at around e.g., $\phi_2 \approx 200$ V at 20 Pa.

At the high SEEC value of $\gamma=0.4$ both $\beta$ and $\overline{\varepsilon}_{\rm e}$ exhibit only a weak dependence on $\phi_2$, because of the significant additional supply of electrons from the electrodes, which are accelerated to high energies and are collisionally multiplied within the sheaths. 

\begin{figure}[ht!]
\begin{center}
\includegraphics[width=0.4\textwidth]{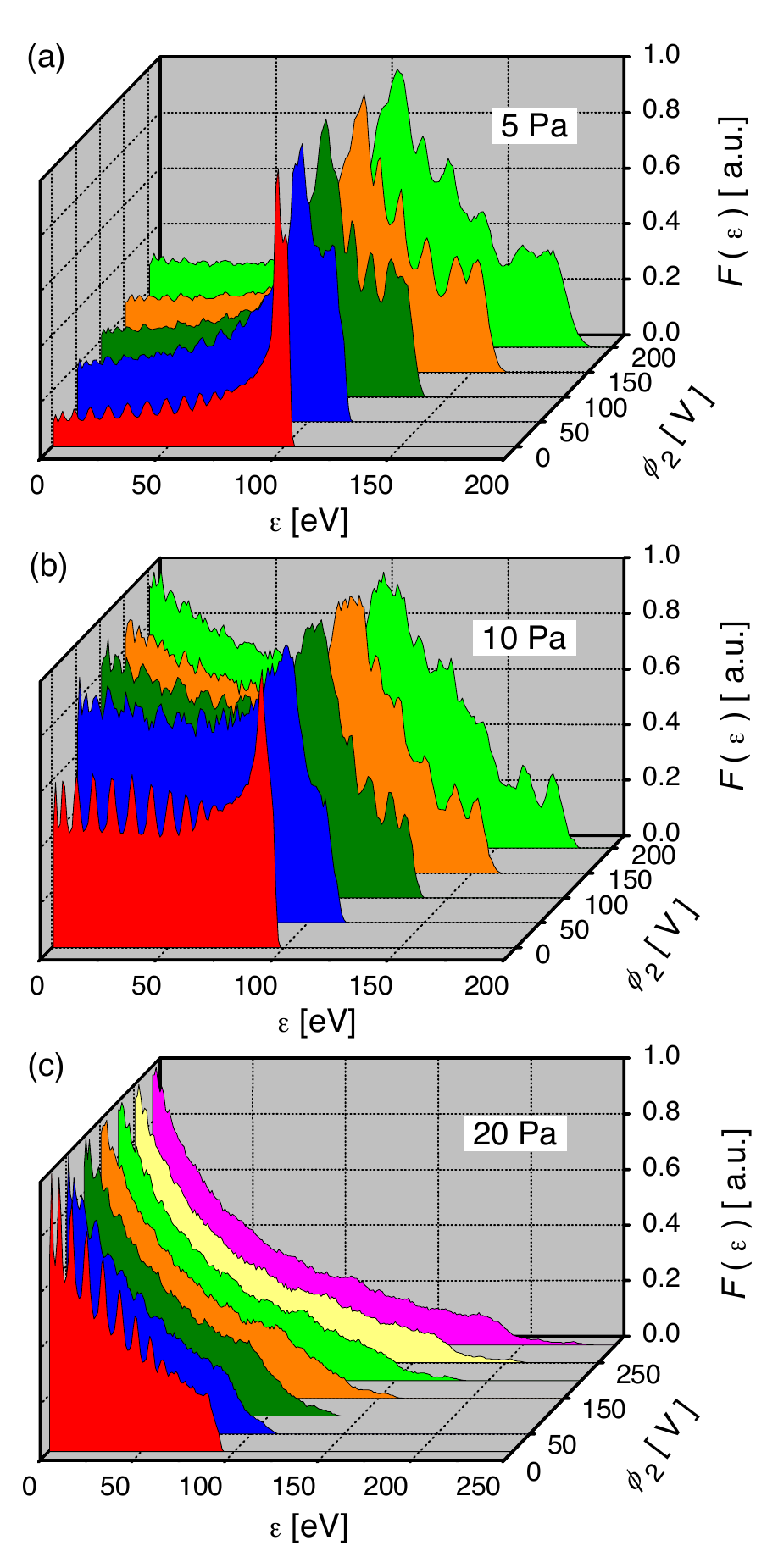}
\caption{Flux-energy distributions of O$_2^+$ ions at the electrodes of oxygen discharges driven by classical dual-frequency waveforms, at different pressures: 5 Pa (a), 10 Pa (b), and 20 Pa (c). Each curve is normalised to a maximum value of 1.0. $\phi_1$ = 200 V and $\gamma=0$  for all cases.}
\label{fig:DF_ifedf}
\end{center}
\end{figure}

The flux-energy distributions of O$_2^+$ ions at the electrodes, as a function of $\phi_2$ are displayed in figure~\ref{fig:DF_ifedf} for the pressures of 5 Pa (a), 10 Pa (b), and 20 Pa (c), for $\gamma=0$. (As the discharge is symmetrical, these distributions are the same at both electrodes.) The results indicate that for each pressure the range of ion energies is expanded as the low-frequency voltage amplitude grows. The influence of $\phi_2$ on the sheath widths (cf. figure \ref{fig:DF_densities}) and consequently, on the collisionality, also influences the shapes of the IFEDFs at low pressures. At 5 Pa pressure, e.g., while the IFEDF is peaked near the maximum energy when $\phi_2$ = 0 V,  the energy spectrum exhibits from $\sim$100 eV onwards a fall with increasing energy, as the ions have little probability to cross the enlarged sheath without charge exchange collisions. At 20 Pa pressure the energy range is expanded without a major change of the shape of the spectrum. A notable difference is, however, the disappearance of the distinct peaks of the IFEDF, which are only present at $\phi_2$~=~0~V. 

The peaks in the IFEDFs can form as a consequence of charge exchange collisions (forming  slow ions) in the region between the electrode and the maximum sheath width. When the electric field is zero for considerable periods of time within this region, these "cold" ions can accumulate and are accelerated together during the next high-field periods. While in a single-frequency discharge (i.e. $\phi_2$ = 0 V) the electric field exhibits zero values within this region for considerable time periods (several tens of nanoseconds), whenever $\phi_2 \neq$ 0 V, the sheath electric field lacks these intervals, making it impossible for the ions to aggregate and be accelerated in a synchronised manner. 

This phenomenon is analysed in more details based on the model of the ion motion in the sheaths, described Schuengel {\it et al.} \cite{Eddi}. This model is based on approximations of the ion density profile in the sheath as given in \cite{Wild}  and of the temporal evolution of the sheath voltage from an equivalent circuit model \cite{EAE-control3}. It allows the calculation of the ion motion in the sheaths and the IFEDFs from the set of input data that consists of the applied voltage waveform (specified for the given case), the maximum sheath width (which is taken as a result of the PIC/MCC simulations and is slightly corrected in the model, for details see \cite{Eddi}), and the ion mean free path (that can be computed from the cross sections). By computing ion trajectories, the energy of the ions upon the arrival at the electrodes as well as the ion transit time can be obtained for ions starting from arbitrary positions. Based on these, the structure of the IFEDFs was successfully explained in \cite{Eddi} for various discharge conditions. Here, we make use of the same model and carry out calculations to explain the characteristic changes of the IFEDFs as an effect of varying $\phi_2$ in classical DF discharges operated in oxygen. 

\begin{figure}[ht!]
\begin{center}
\includegraphics[width=0.62\textwidth]{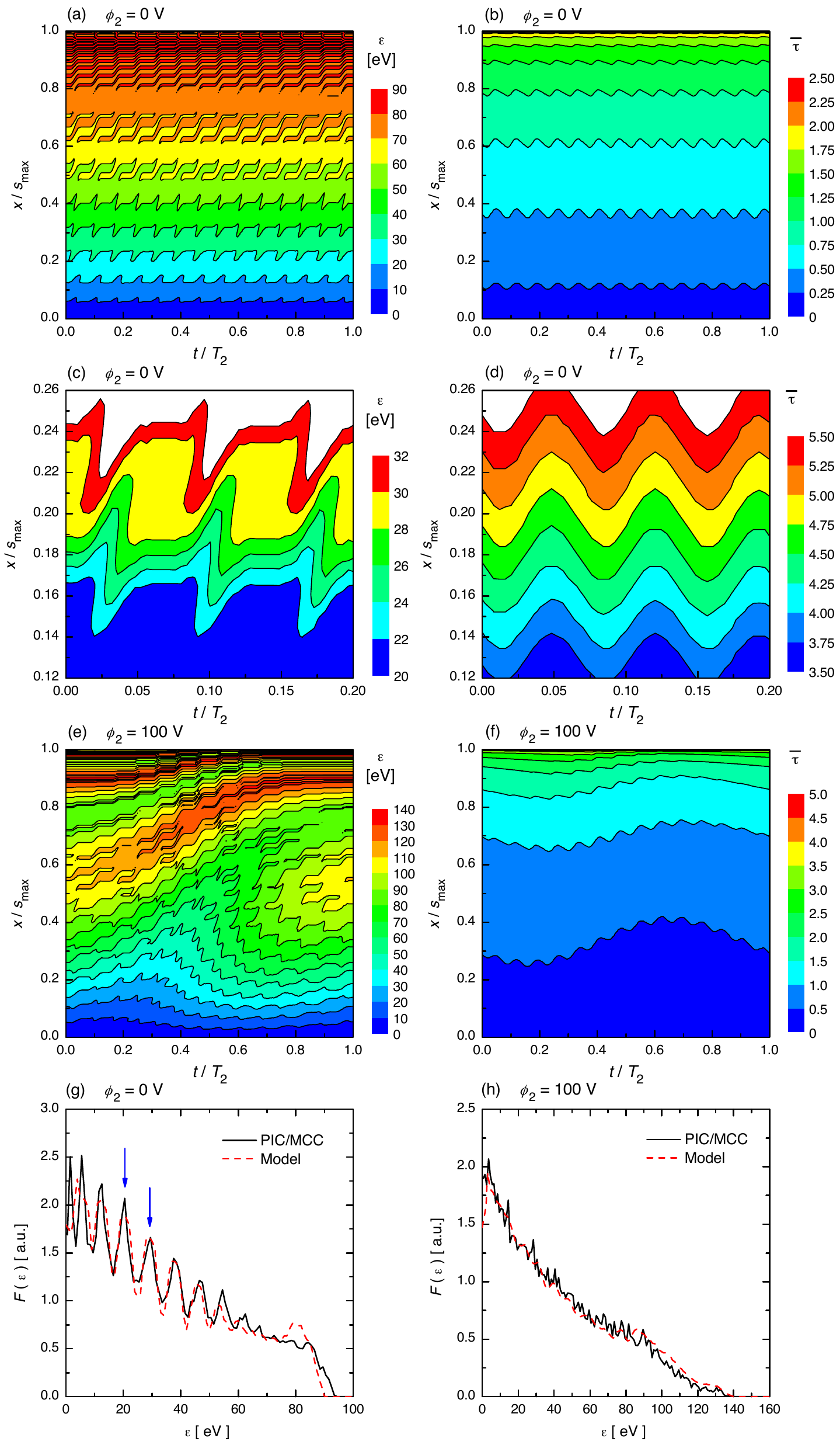}
\caption{(a) Energy of those ions arriving at the electrode, which experienced a charge exchange collision within the sheath at a normalised position $x/s_{\rm max}$ and at a normalised time $t/T_2$ and arrive at the electrode without further collisions and (b) the time that these ions need to arrive at the electrode, $\overline{\tau}  = t / T_1$, for $\phi_2$ = 0 V. (c) and (d) are zoomed regions of (a) and (b). (e) and (f) are the same as (a) and (b), but for $\phi_2$~=~100~V. (g) and (h) show the IFEDFs obtained from the model in comparison with the PIC/MCC simulation results for $\phi_2$ = 0 V and $\phi_2$ = 100 V, respectively. $s_{\rm max}$ is the maximum sheath width, $T_1$ is the period of the high-frequency component and $T_2$ is the period of the low-frequency component of the excitation waveform. $\phi_1$~=~200~V and $p$ = 20 Pa for all cases.}
\label{fig:DF_EDDI}
\end{center}
\end{figure}

\begin{figure}[ht!]
\begin{center}
\includegraphics[width=0.4\textwidth]{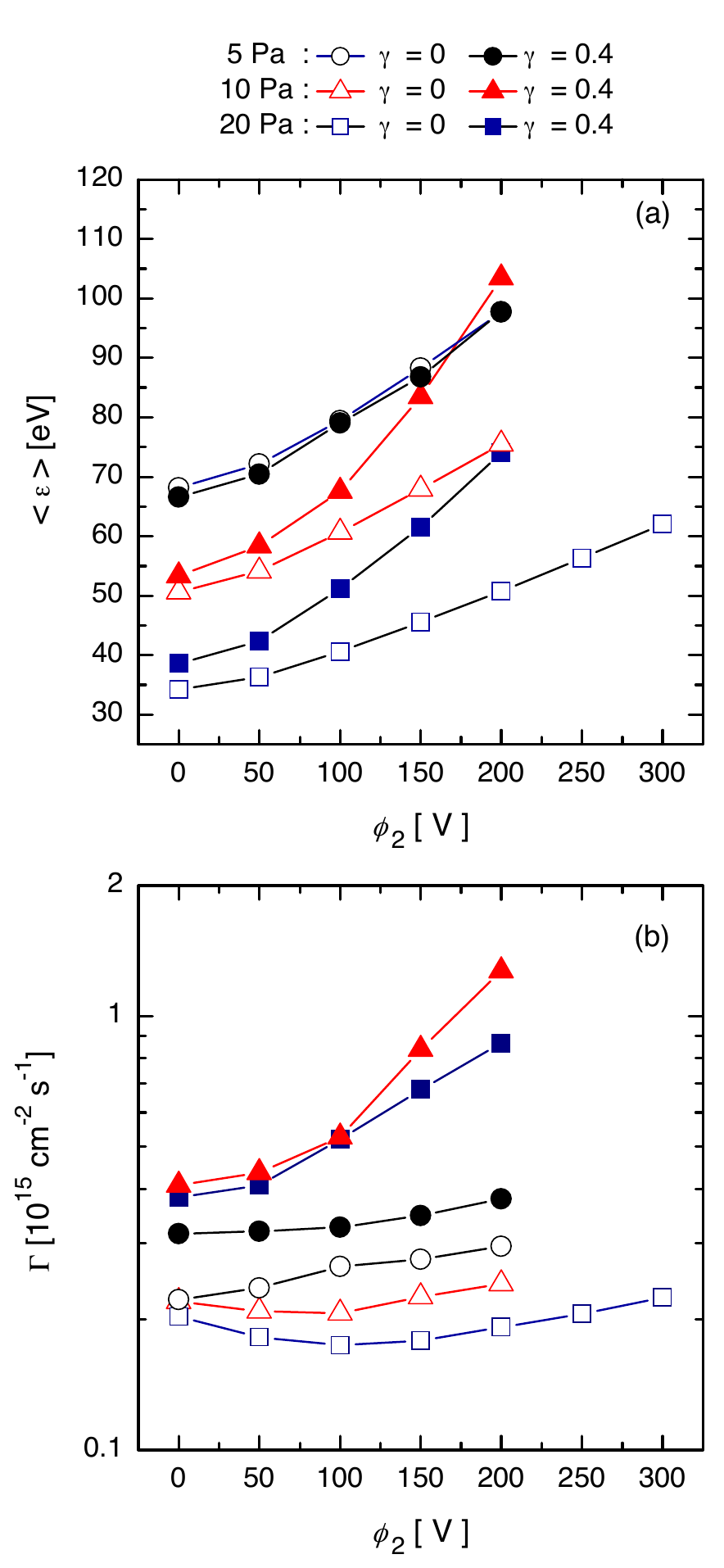}
\caption{Mean energy, $\langle \varepsilon \rangle$, (a) and flux, $\Gamma$, (b) of O$_2^+$ ions at the electrodes, as a function of the low-frequency voltage amplitude, at different pressure and $\gamma$ values specified. $\phi_1$ = 200 V for all cases.}
\label{fig:DF_results}
\end{center}
\end{figure}

Figure \ref{fig:DF_EDDI}(a) shows the energy of those ions arriving at the electrode, which experienced a charge exchange collision within the sheath at a normalised position $x/s_{\rm max}$ (where $s_{\rm max}$ is the maximum sheath width) and at a normalised time $t/T_2$ (where $T_2$ is the period of the low-frequency component of the excitation waveform) and arrive at the electrode without further collisions, in the case $\phi_2$ = 0 V. Panel (b) shows the time that these ions need to arrive at the electrode, $\overline{\tau}$, that is normalised by $T_1$ (where $T_1$ is the period of the high-frequency component of the excitation waveform). These plots, as well as panels (c) and (d), which are zoomed parts of panels (a) and (b), confined to narrower domains of space and time, indicate that there are certain zones in space and time, where ions, that have undergone a charge exchange collision, can accumulate. For these domains the same ion energy at the electrode is achieved. As an example, the peaks in the IFEDF shown in figure \ref{fig:DF_EDDI}(g) at about 21 eV and 29 eV (marked with arrows in the plot) originate from ions that were "born" in $x/s_{\rm max}$ of about 0.14 and 0.22, respectively (blue and yellow regions in panel (c)). These ions need 4 and 5 periods of the 27.12 MHz cycle, respectively, to reach the electrode (as inferred from figure \ref{fig:DF_EDDI}(d). The IFEDF constructed from the model this way (see figure \ref{fig:DF_EDDI}(g)) matches extremely well the distribution obtained from the PIC/MCC simulation.

The energy of the arriving ions and the time needed to reach the electrode following a charge exchange collision is displayed, respectively, in figures \ref{fig:DF_EDDI}(e) and (f), for $\phi_2$ = 100 V, while the resulting IFEDF is depicted in figure \ref{fig:DF_EDDI}(h). At these conditions there are no clear zones in space and time, where ions that have undergone a charge exchange collision may accumulate. Ions that are born at the same spatial position but at different times are accelerated differently due to the presence of the low-frequency field. Therefore, they reach the electrode with different energies. The model again predicts an IFEDF that is in very good agreement with the one obtained from the PIC/MCC simulation, and explains the changes of the IFEDF due to a change of $\phi_2$ in an elementary way.

According to figure \ref{fig:DF_ifedf} the effect of $\phi_2$ on the IFEDFs is appreciable at all pressures. This shows the possibility of the control of the mean ion energy by the $\phi_2$ amplitude. This voltage amplitude should, in principle, have a negligible effect on the flux of the ions. Previous studies of the independent control of ion properties have, however, shown that this is violated due to (i) frequency coupling effects and (ii) the influence of secondary electron emission from the electrodes. The first effect results from the fact that the sheath dynamics is determined by both voltage amplitudes and as an increasing low-frequency voltage expands the sheaths, in general, the high-frequency sheath width oscillations take place in the domain of higher ion densities, thereby decreasing the modulation of the sheath width and the speed of expansion \cite{coupling1,coupling2}. The second effect (the influence of secondary electrons) results from the higher total accelerating voltage which may give rise to significantly higher multiplication of secondary electrons creating a higher plasma density and higher ion flux \cite{secondaries1,secondaries2}. The results of our investigation of the performance of the separate control of the mean ion energy and ion flux are presented in figure~\ref{fig:DF_results}. Panel (a) shows the mean ion energy, $\langle \varepsilon \rangle$, while panel (b) shows the  O$_2^+$ ion flux, $\Gamma$, as a function of the low-frequency voltage amplitude. As the discharge is symmetrical, these quantities are the same at both electrodes. The data reveal that the value of $\phi_2$ controls well the mean ion energy in all cases, however, for some cases this is accompanied by a change of the ion flux. Figure~\ref{fig:DF_results}(b) shows that this effect is most critical at the highest value of the SEEC, $\gamma=0.4$. In this case the independent control is impossible at the higher pressures (10 Pa and 20 Pa). In the other parameter combinations the flux remains reasonably constant, providing a way of controlling the mean ion energy (closely) independently of the ion flux $\Gamma$.

\subsection{Valleys-waveforms}

\label{sec:valleys}

In the following we present simulation results for multi-frequency waveforms, defined by eq.\,(\ref{eq:exc3}), with a base frequency of $f_1$ = 15 MHz and including harmonics up to $N$ = 4. This type of waveform was found in previous studies to be efficient for an independent control of the ion properties \cite{Derzsi2016,EAE-control1,EAE-control2,EAE-control3}. The basis of this control is the self-bias voltage that develops (even in geometrically symmetrical reactors) due to the amplitude asymmetry of this waveform. The control parameters, which determine the value of the DC self-bias (at fixed $\phi_k$ harmonic amplitudes) are the $\theta_k$ phase angles in eq.\,(\ref{eq:exc3}). As mentioned in section 2.1, the value of the DC self-bias voltage is determined in an iterative manner in the simulations, to ensure equal losses of positive and negative charges at each electrode over one period of the fundamental driving frequency \cite{DZ-EAE}, i.e. the self-bias voltage is not pre-defined for the given waveform, but it is self-consistently calculated.

Below, examples will be given for valleys-type waveforms that give rise to a positive self-bias. Therefore, the ion energy range is expected to be extended at the grounded electrode due to the increase of the time-averaged sheath voltage drop at that side of the plasma. In contrast, a decrease of the mean ion energy at the powered electrode is contemplated with increasing self-bias. 

We do not carry out a variation of the phase angles, the presentation of the results is restricted to phases $\theta_k =0$ for each odd harmonic and $\theta_k =\pi$ for each even harmonic, specific for {\it valleys-type} waveforms. The possibility of the control of the mean ion energy is, however, still revealed at this choice of the phase angles, as the self-bias voltage takes its extremum value near $\theta_k =0$ \cite{DZ-EAE}. Thus, the mean ion energies at the two electrodes are (very nearly) maximum and minimum values, i.e. the range between them represents the control range for the mean ion energy. 

\begin{figure}[ht!]
\begin{center}
\includegraphics[width=0.4\textwidth]{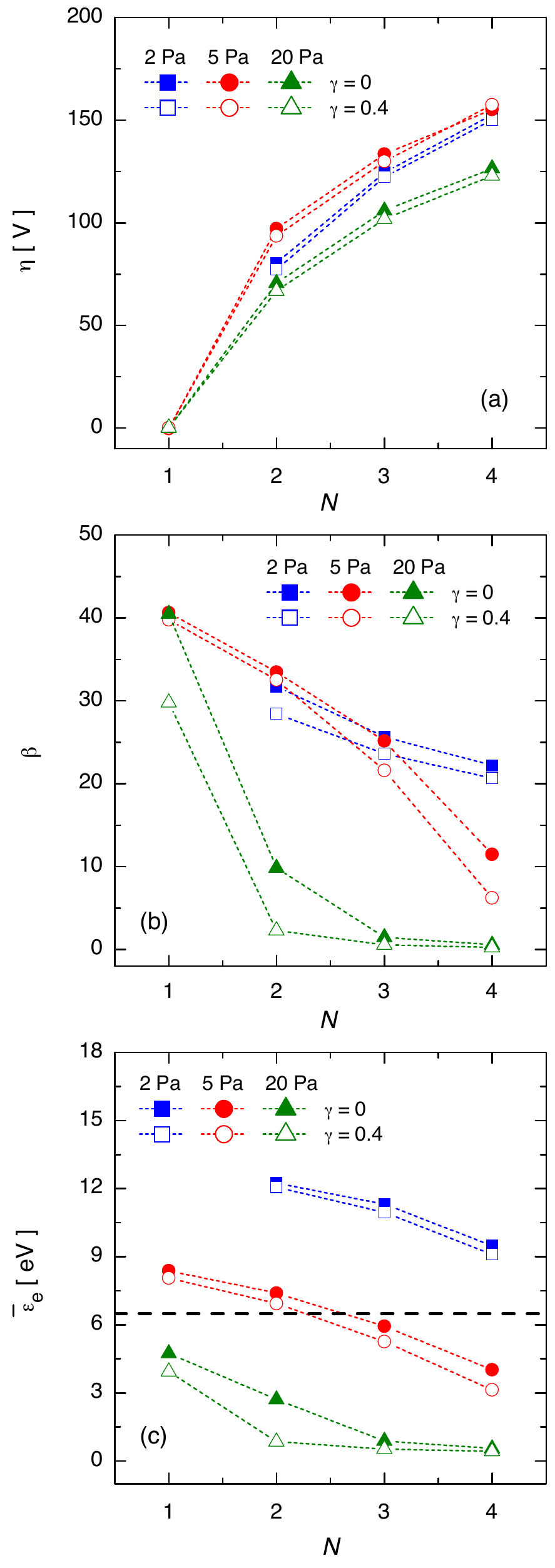}
\caption{Self-bias voltage, $\eta$, (a), electronegativity, $\beta$, (b), and (temporally averaged) mean electron energy, $\overline{\varepsilon}_{\rm e}$, in the discharge centre (c) in oxygen discharges driven by valleys-type waveforms (specified by eq.\,(\ref{eq:exc3})) consisting of $N$ harmonics, at different pressures and $\gamma$ values. The dashed horizontal line in (c) indicates the energy (6.5 eV) where the attachment cross section peaks.  $\phi_{\rm pp}$ = 400 V for all cases.}
\label{fig:valleys_bias}
\end{center}
\end{figure}

\begin{figure}[ht!]
\begin{center}
\includegraphics[width=0.4\textwidth]{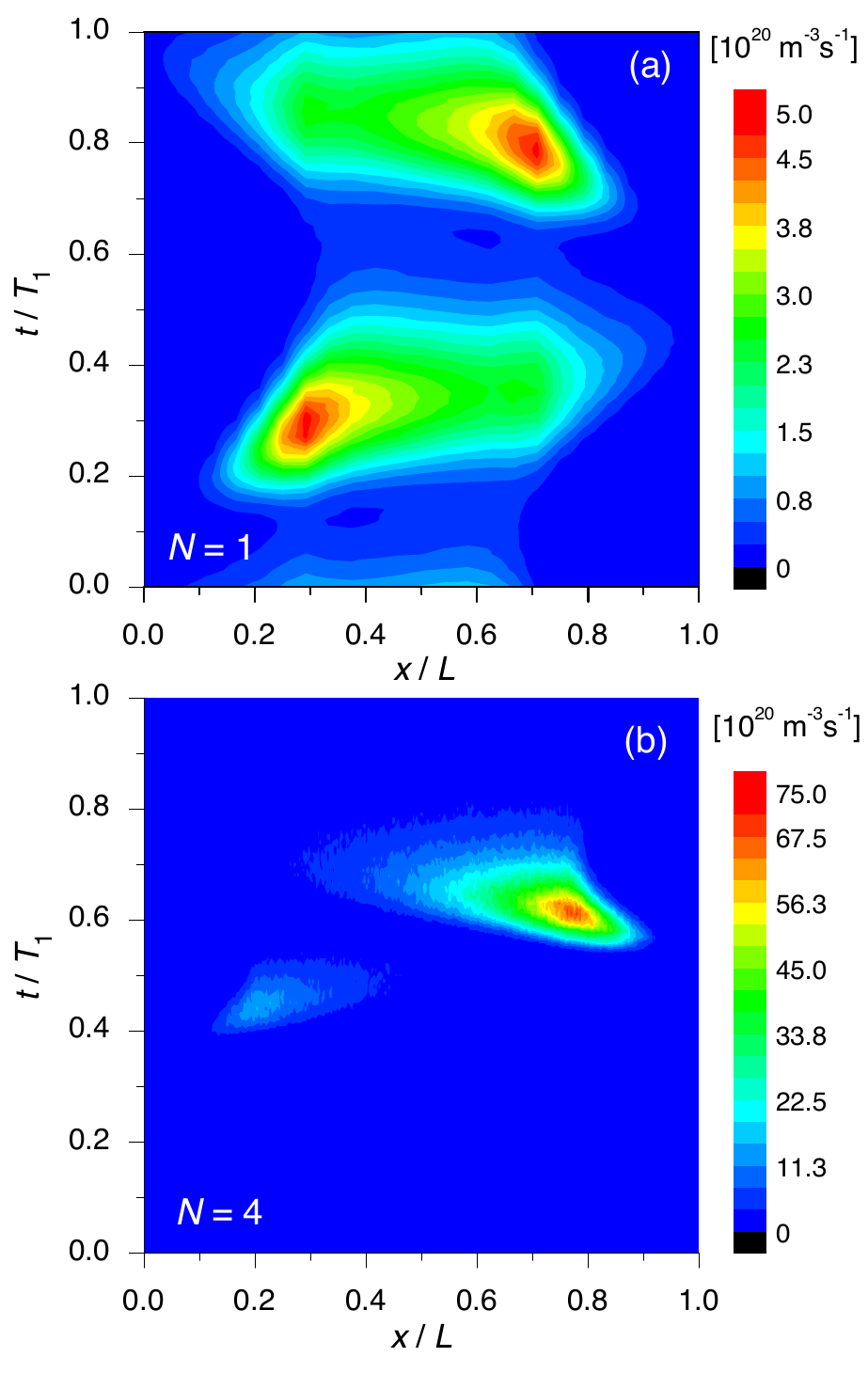}
\caption{Spatio-temporal distribution of the ionisation rate in oxygen discharges driven by valleys-type waveforms (specified by eq.\,(\ref{eq:exc3})) consisting of $N=1$ (a) and $N=4$ (b) harmonics. $\phi_{\rm pp}$ = 400 V, $p$ = 20 Pa and $\gamma=0$ for both cases.  $T_1$ is the period of the fundamental frequency ($f_1$).}
\label{fig:valeys_ion_rate}
\end{center}
\end{figure}

First, we present the PIC/MCC simulation results for the computed DC self-bias voltage ($\eta$) and the electronegativity ($\beta$) of the plasma, in figure \ref{fig:valleys_bias}(a) and (b), respectively. The data shown in figure \ref{fig:valleys_bias}(a) confirm that a significant DC self-bias voltage is established when the waveform includes an increasing number of harmonics. Accompanied by this change is a decrease of the electronegativity, as seen in figure \ref{fig:valleys_bias}(b). The latter effect is most pronounced at 20 Pa. The change of $\beta$ strongly correlates with the change of the (temporally averaged) mean electron energy at the discharge centre, shown in figure \ref{fig:valleys_bias}(c). In the case of valleys waveforms the observed decrease of the mean electron energy ($\overline{\varepsilon}_{\rm e}$) as a function of $N$ is caused by an electron power absorption mode transition, as shown in figure \ref{fig:valeys_ion_rate} that compares the spatio-temporal distribution of the ionisation rates obtained for $N=1$ and $N=4$ (at $\phi_{\rm pp}$ = 400 V, $p$ = 20 Pa and $\gamma=0$). At $N=1$ a hybrid $\alpha$ + DA power absorption mode is revealed, with significant ionisation within the plasma bulk region. This is the consequence of the high electric field within that domain under the conditions of high electronegativity (see figure \ref{fig:valleys_bias}(b)) i.e., strongly depleted electron density. When $N$ is changed to 4, a pure $\alpha$-mode electron power absorption establishes, ionisation is concentrated near the expanding sheath edges, with a strong dominance of the region near the grounded electrode, due to the strong positive self-bias at these conditions (see figure \ref{fig:valleys_bias}(a)). The data indicate a $\approx$~15 times higher peak ionisation rate in the case of $N$ = 4, compared to the case of $N$ = 1, while the spatio-temporal average of these also shows an increase by a factor of $\approx$~2.5. Increasing $N$ enhances the sheath expansion heating of electrons. This corresponds to an enhanced source of energetic electrons, which explains the decrease of the mean electron energy observed in figure \ref{fig:valleys_bias}(c). Similar to the classical dual-frequency scenario, a change of the mean electron energy is coupled to a change of the electronegativity ($\beta$) via the energy dependence of the attachment cross section (see section \ref{sec:dual}).

\begin{figure}[ht!]
\begin{center}
\includegraphics[width=0.4\textwidth]{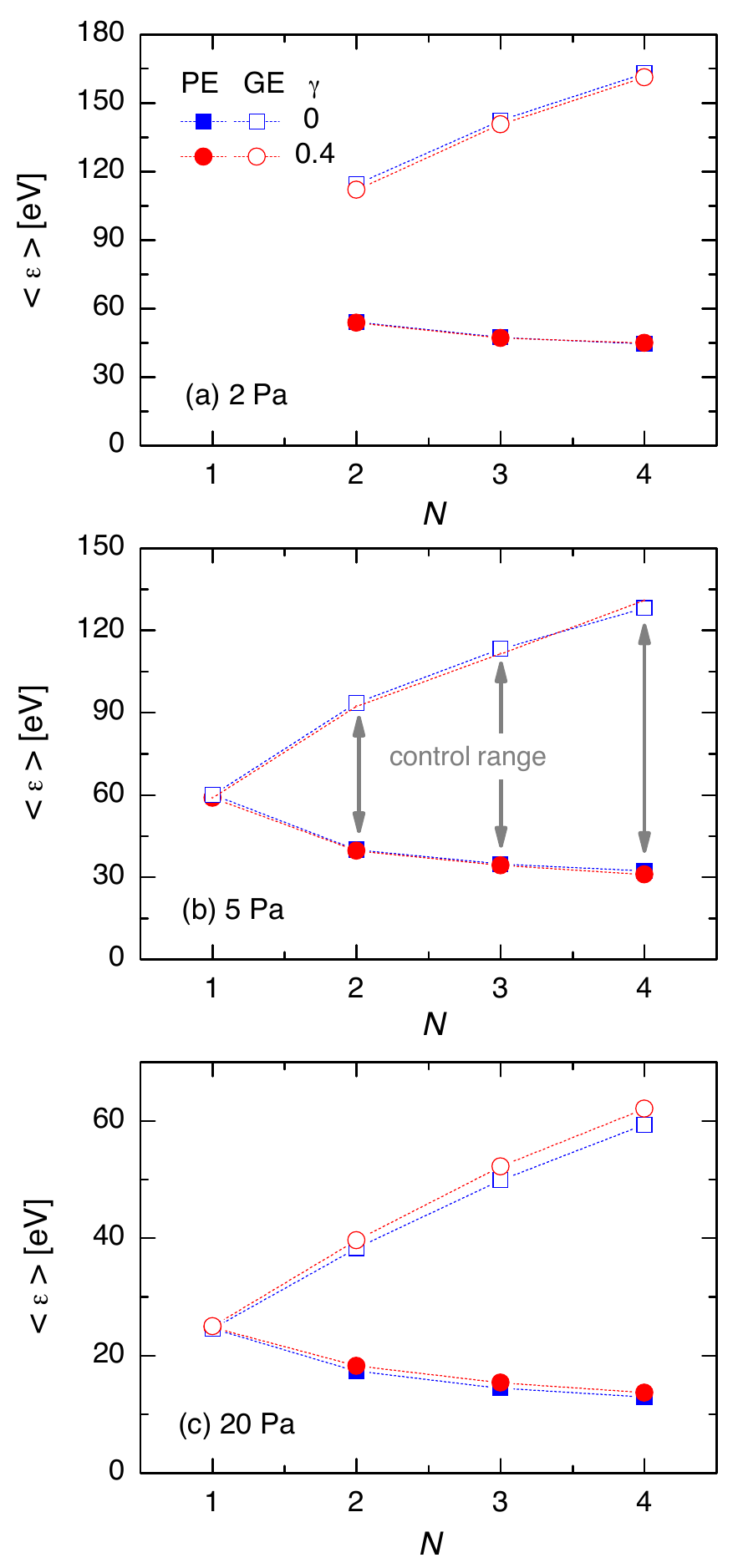}
\caption{Mean energy of O$_2^+$ ions at the powered electroed (PE) and at the grounded electrode (GE) of oxygen discharges driven by valleys-type waveforms (specified by eq.\,(\ref{eq:exc3})) consisting of $N$ harmonics, at different pressures. $\phi_{\rm pp}$ = 400 V and $\gamma=0$ for all cases. The arrows in the case of 5 Pa show the control range of the mean ion energy that can be covered by changing the phase angles in the driving voltage waveform. }
\label{fig:valleys_flux_energy}
\end{center}
\end{figure}

The change of the DC self-bias with increasing number of harmonics ($N$) is expected to change the mean energy of ions arriving at the electrodes. This effect is confirmed by the results shown for the mean ion energy in figure \ref{fig:valleys_flux_energy}. We note that at 2 Pa pressure it was not possible to ignite the plasma with a single harmonic, $N=1$. Taking the $p$~=~5~Pa case as an example, the data show that using the highest number of harmonics, the maximum and minimum mean ion energies are 130 eV and 30 eV, respectively, i.e., $\langle \varepsilon \rangle$ can be changed by more than a factor of 4 at the same conditions by changing the phase angles of the driving voltage waveform. The mean ion energy is insensitive to the value of $\gamma$.

\begin{figure}[ht!]
\begin{center}
\includegraphics[width=0.4\textwidth]{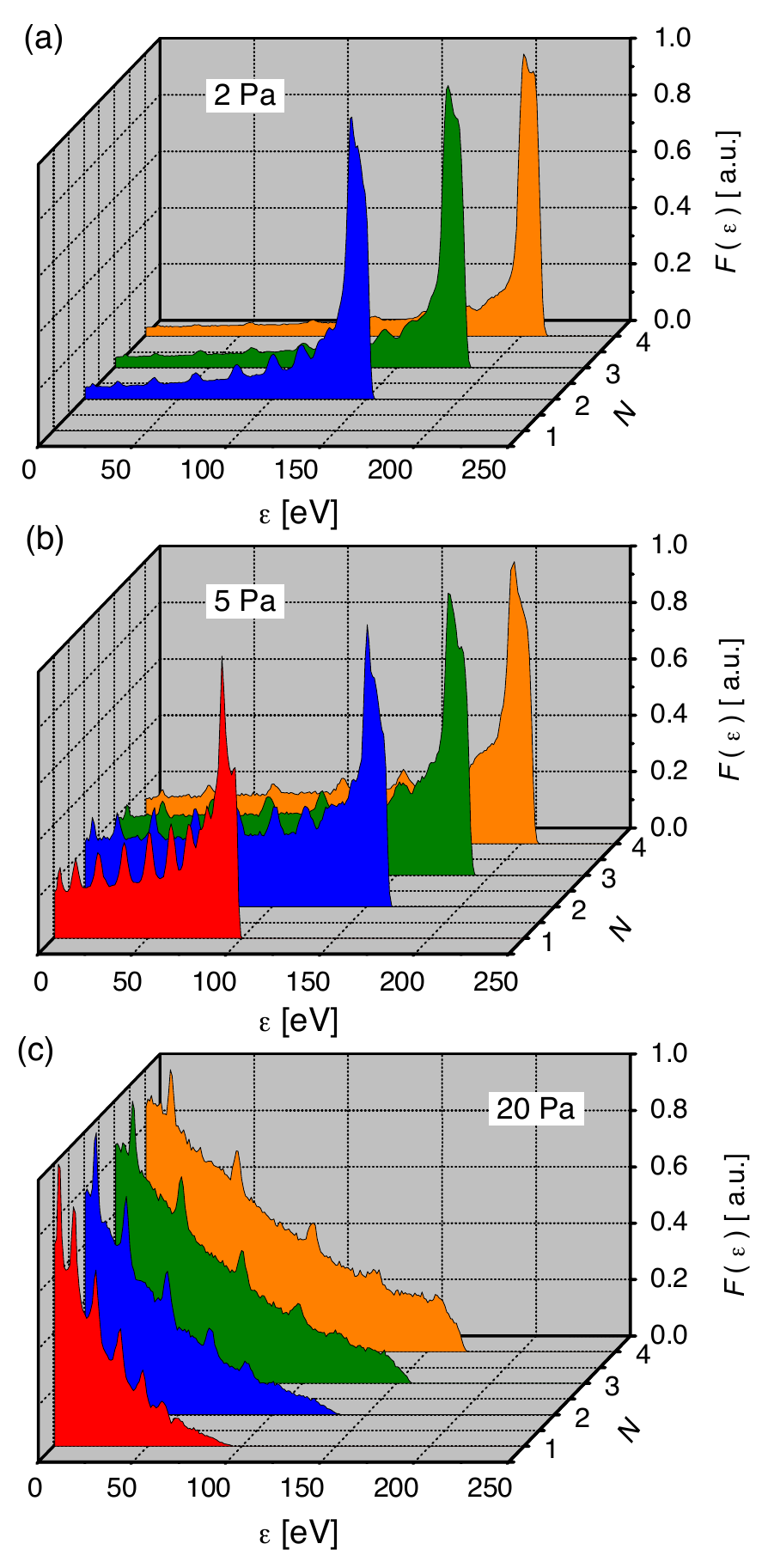}
\caption{Flux-energy distributions of O$_2^+$ ions at the grounded electrode of oxygen discharges driven by valleys-type waveforms (specified by eq.\,(\ref{eq:exc3})) consisting of $N$ harmonics, obtained at different pressures: 2 Pa (a), 5 Pa (b), and 20 Pa (c). Each curve is normalised to a maximum value of 1.0. $\phi_{\rm pp}$ = 400 V and $\gamma=0$.}
\label{fig:valleys_iedfs}
\end{center}
\end{figure}

\begin{figure}[ht!]
\begin{center}
\includegraphics[width=0.4\textwidth]{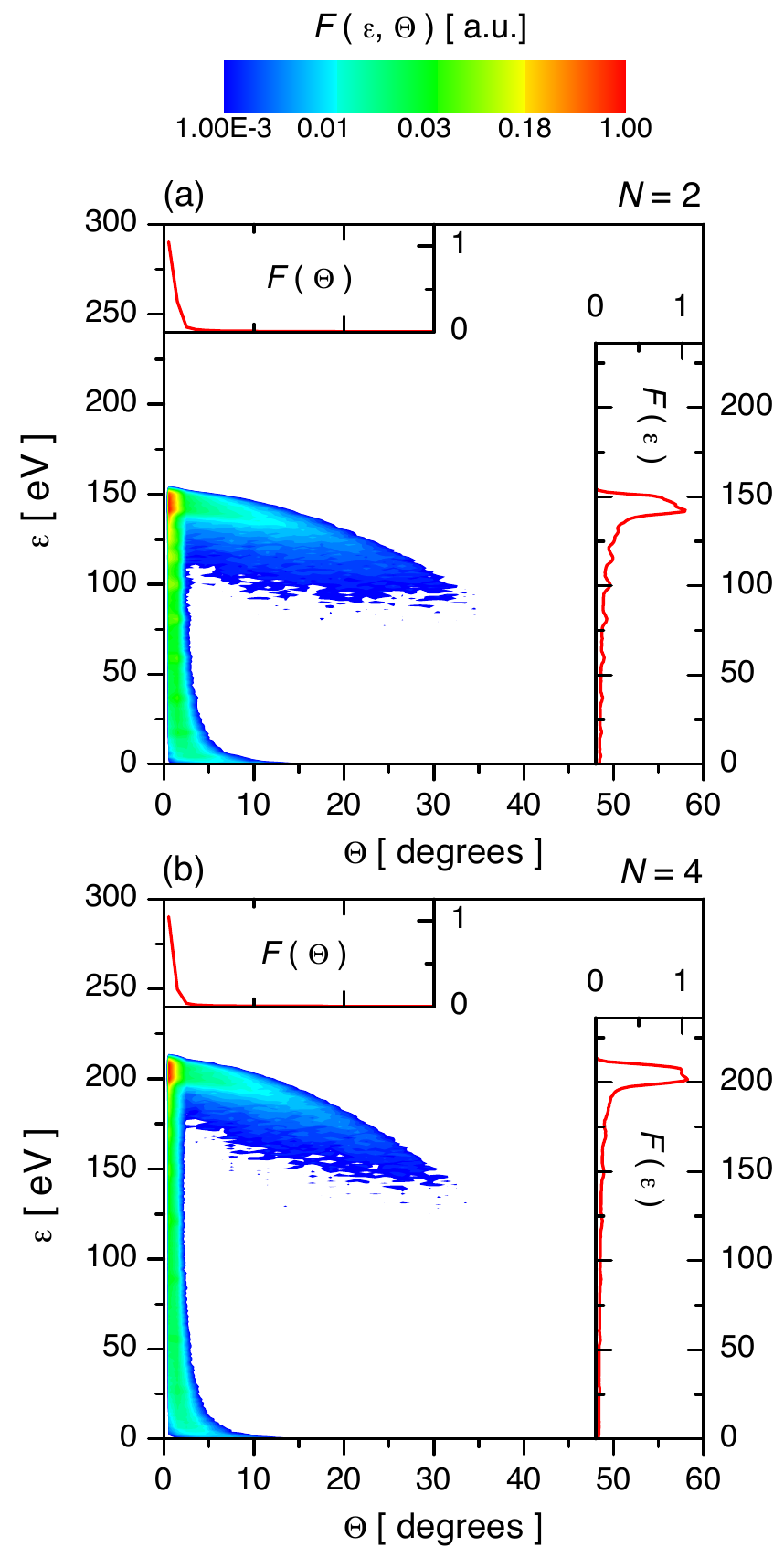}
\caption{Energy and angular distribution of the O$_2^+$ ion flux, $F(\varepsilon,\Theta)$ (in arbitrary units) at the grounded electrode in discharges driven by valleys-type waveforms with $N$ = 2 (a) and $N$ = 4 (b) harmonics, at 2 Pa pressure. The insets illustrate the energy and angular distributions (respective integrals of $F(\varepsilon,\Theta)$ according to incidence angle and energy). All distributions have been normalised to a maximum value of 1, for easier comparison. $\phi_{\rm pp}$ = 400 V, $\gamma=0$.}
\label{fig:valleys_angular}
\end{center}
\end{figure}

The flux-energy distribution function of the O$_2^+$ ions at the grounded electrode of the CCP are shown in figure \ref{fig:valleys_iedfs} for different pressures, as a function of the number of harmonics in the valleys-type excitation waveform. The range of ion energy is significantly extended at all pressures when the number of applied harmonics is increased. We observe, however, a significant difference of the shape of the IFEDF compared to that obtained for the classical dual-frequency excitation, compare, e.g., the 5 Pa cases  (figure~\ref{fig:DF_ifedf}(a) vs. figure~\ref{fig:valleys_iedfs}(b)). This difference originates from the fact that while in the case of the classical DF excitation the low-frequency voltage strongly increases the sheath width, this is not the case when the valleys-type waveform is applied. Therefore, these two types of excitation result in noteworthily different IFEDFs. Enhancing the ion energies without increasing the collisionality of the sheaths is definitely advantageous when a well-directed beam of ions is necessary for surface processing. The valleys-type (or peaks-type) waveforms provide this possibility, as figure \ref{fig:valleys_iedfs} confirms. For the lowest pressure case a nearly mono-energetic beam of ions arrives at the electrode for each $N$. More details about these cases (at 2 Pa pressure) are given in figure~\ref{fig:valleys_angular}, which shows the $F(\varepsilon,\Theta)$ combined, energy and angularly-resolved distribution function.

\begin{figure}[ht!]
\begin{center}
\includegraphics[width=0.4\textwidth]{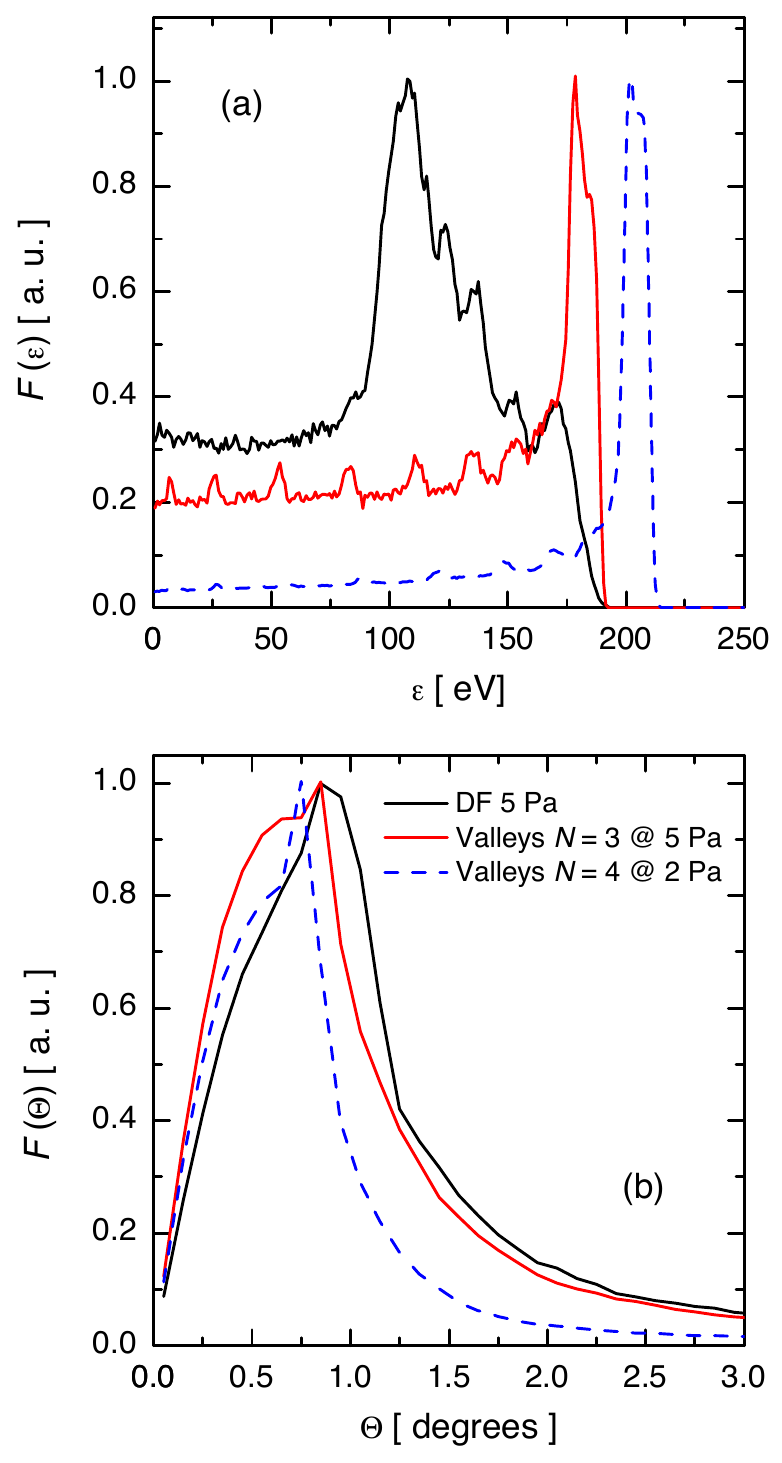}
\caption{Comparison of flux energy (a) and angular (b) distributions of the O$_2^+$ ions (in arbitrary units) at the grounded electrode in discharges driven by a DF waveform ($\phi_1 = \phi_2$ = 200 V) at 5 Pa, a valleys-waveform with $N$ = 3 at 5 Pa, and a valleys-waveform with $N$ = 4 at 2 Pa. These distributions have been normalised to peak values of 1, for an easier comparison of their {\it shapes}. $\gamma=0$.}
\label{fig:comp_angular}
\end{center}
\end{figure}

Figure \ref{fig:comp_angular} compares the IFEDFs and IADFs obtained with DF and valleys waveforms. The DF excitation at 5 Pa and $\phi_1 = \phi_2$ = 200 V and the valleys-type excitation with 3 harmonics, at 5 Pa and $\phi_{\rm pp}$ = 400 V, form a pair of cases in the sense that the maximum ion energy is very nearly the same. Significantly different flux-energy distributions are obtained for these cases as already mentioned above. The difference between the sheath widths in these two cases, that causes the different shapes of the IFEDFs, also results in different angular distributions of the arriving ions; in the case of valleys-type excitation the distribution shifts to lower values of $\Theta$. A further narrowing of the $F(\Theta)$ distribution can be achieved by operating the plasma at a lower pressure, with a higher number of harmonics, as it is illustrated in figure \ref{fig:comp_angular} for valleys-type excitation with $N$ = 4 harmonics, at 2 Pa and $\phi_{\rm pp}$ = 400 V.

\clearpage

\subsection{Sawtooth-waveforms}

\begin{figure}[h!]
\begin{center}
\includegraphics[width=0.38\textwidth]{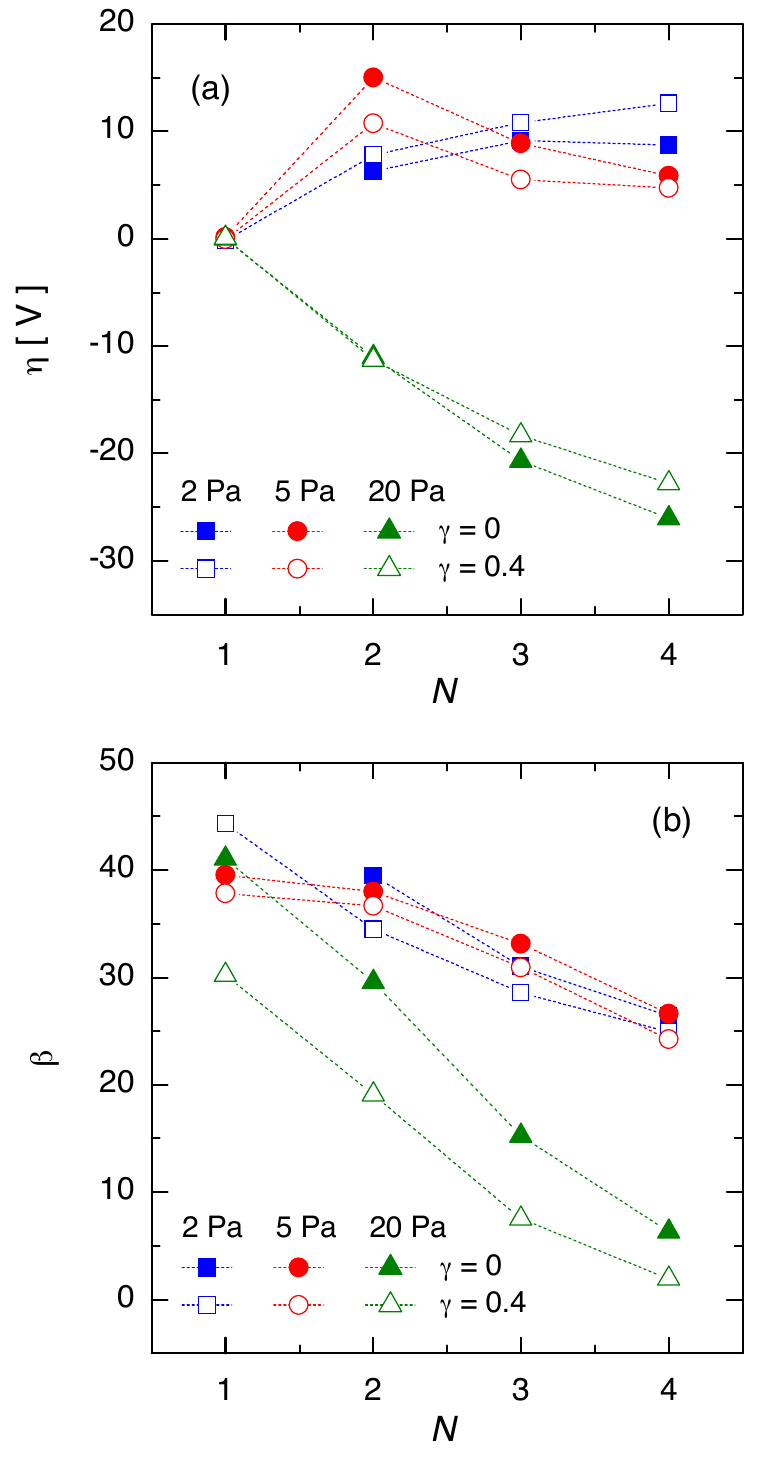}
\caption{Self-bias voltage, $\eta$, (a) and electronegativity, $\beta$, (b) in oxygen discharges driven by sawtooth-type waveforms (specified by eq.\,(\ref{eq:exc4})) consisting of $N$ harmonics, at different pressures. $\phi_{\rm pp}$ = 400 V for all cases.}
\label{fig:sawtooth_bias}
\end{center}
\end{figure}

\begin{figure}[h!]
\begin{center}
\includegraphics[width=0.42\textwidth]{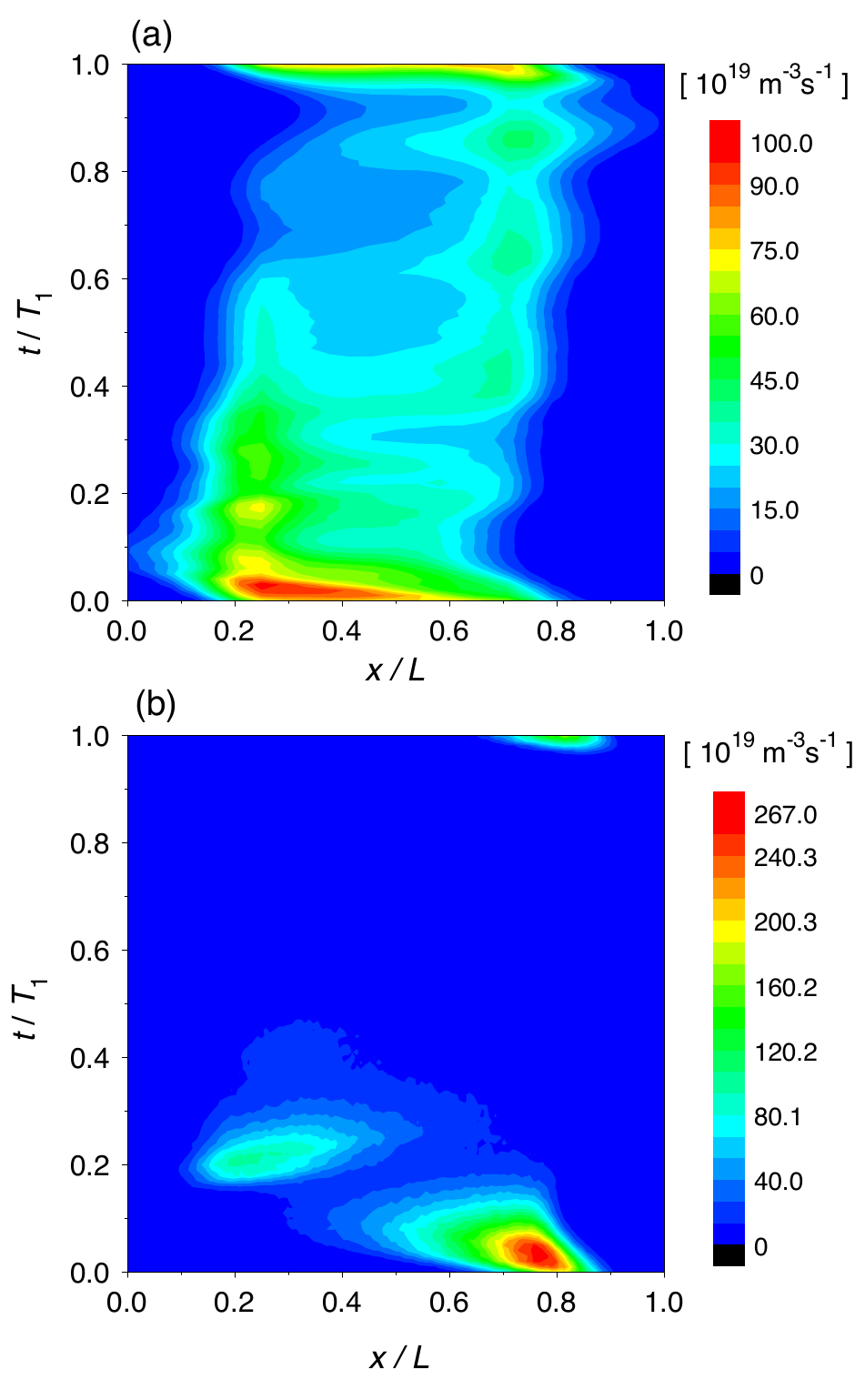}
\caption{Spatio-temporal distribution of the ionisation rate in oxygen discharges driven by sawtooth-down waveforms (specified by eq.\,(\ref{eq:exc4})) consisting of $N$ = 4 harmonics, for $p$ = 2 Pa (a) and $p$ = 20 Pa (b) pressures. $T_1$ is the period of the fundamental frequency ($f_1$). $\phi_{\rm pp}$ = 400 V and $\gamma=0$ for both cases.}
\label{fig:sawtooth_ionisation}
\end{center}
\end{figure}

\begin{figure}[h!]
\begin{center}
\includegraphics[width=0.4\textwidth]{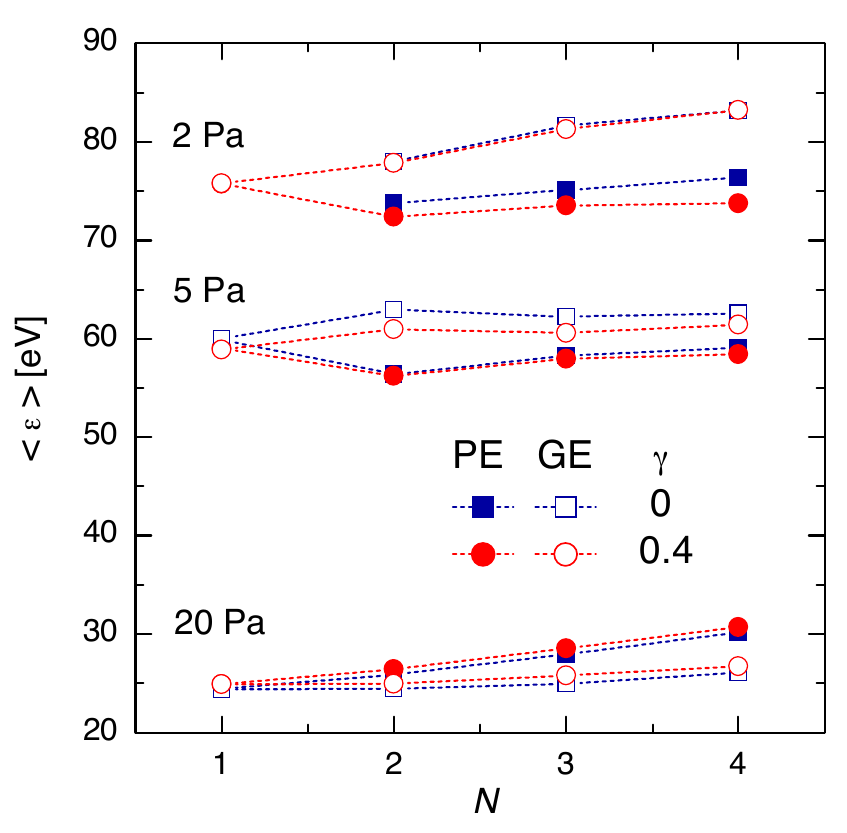}
\caption{Mean energy of the O$_2^+$ ions at the powered electrode (PE, filled symbols) and at the grounded electrode (GE, open symbols) of oxygen discharges driven by sawtooth-down waveforms (specified by eq.\,(\ref{eq:exc4})) consisting of $N$ harmonics, at different pressures. $\phi_{\rm pp}$ = 400 V and $\gamma=0$ for all cases.}
\label{fig:sawtooth_energy_flux}
\end{center}
\end{figure}

Finally, we address the properties of CCPs driven by sawtooth-waveforms. We consider {\it sawtooth-down} waveforms resulting from a plus sign in eq.\,(\ref{eq:exc4}). Unlike in the case of valleys-type (or peaks-type) waveforms that exhibit different positive and negative extrema ("amplitude asymmetry effect"), the sawtooth-type waveforms have equal positive and negative extrema. The difference of the rise and fall times of the waveform, however, still establishes an asymmetry of the discharge dynamics ("slope asymmetry") that can give rise to a self-bias voltage, different sheath properties and IFEDFs at the two electrodes, spatial distributions of the excitation rates, etc. (The asymmetry of the discharge depends strongly on the type of the buffer gas used; discharges in Ar, CF$_4$, and H$_2$ gases have been studied and completely different excitation dynamics were found at the same driving voltage waveforms  \cite{Bastien,Bastien2}.) In the present study, the number of harmonics ranges between $N=1$ and $N$ = 4. We find that the behaviour of the self-bias for oxygen gas is rather specific, as it can be seen in figure~\ref{fig:sawtooth_bias}(a). At low pressures (2 Pa and 5 Pa) a small positive self-bias voltage is generated when the number of harmonics is increased. In contrast with this, a negative self-bias appears at the high pressure (20 Pa) case, with a significantly higher magnitude compared to that in the low-pressure cases. The electronegativity of the plasma decreases with increasing number of harmonics, similarly to the behaviour observed in the case of the valleys-type waveforms. The strongest dependence is also observed here at the highest pressure. These changes can be understood by analysing the spatio-temporal distributions of the ionisation rate, which are shown in figure \ref{fig:sawtooth_ionisation} for $p$ = 2 Pa and $p$ = 20 Pa pressures, at $N$ = 4 harmonics, $\phi_{\rm pp}$ = 400 V and $\gamma=0$. The change of the gas pressure induces a change of the electron power absorption mode. At the lower pressure the highest rate of ionisation appears in the bulk plasma, while at the higher pressure it is confined within a narrow $x-t$ region, near the edge of the expanding sheath. Similar to the observations of Bruneau {\it et al.} \cite{Bastien2} this changes the symmetry of the discharge and therefore the sign of the self-bias. We note that the present pressure dependence is opposite to that observed in \cite{Bastien2} for CF$_4$ gas, because O$_2$ is electronegative at low pressure and CF$_4$ is electronegative at high pressure.

\begin{figure}[h!]
\begin{center}
\includegraphics[width=0.4\textwidth]{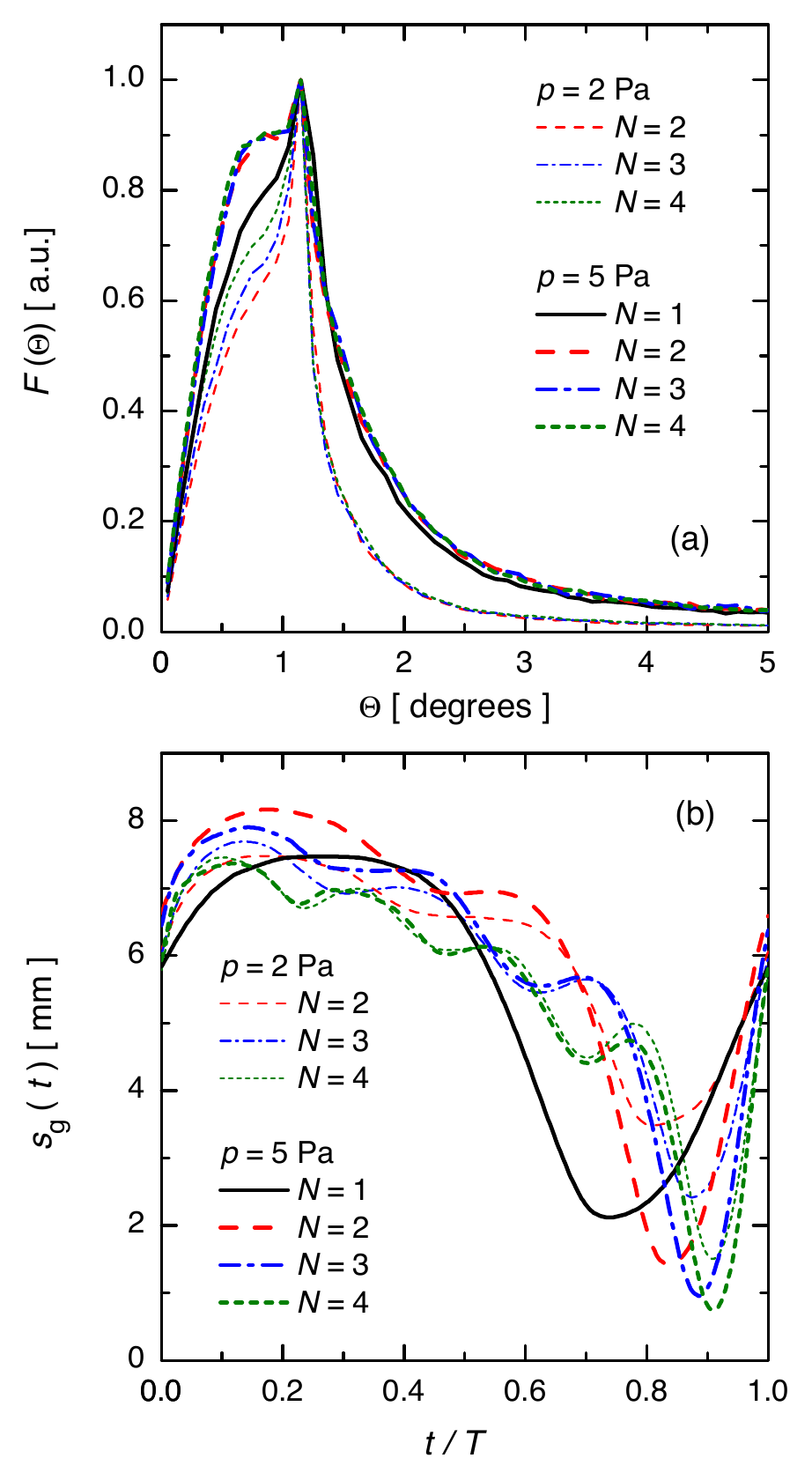}
\caption{(a) Angular distribution of the O$_2^+$ ions at the grounded electrode in discharges driven by sawtooth-down waveforms with different number of harmonics ($N$), at pressures of $p$ = 2 Pa (thin lines) and 5 Pa (thick lines). The distributions have been normalised to a maximum value of 1, for easier comparison. (b) Time dependence of the sheath width at the grounded electrode for the same cases as shown in (a).  $\phi_{\rm pp}$~=~400~V and $\gamma$~=~ 0.}
\label{fig:sawtooth_angular0}
\end{center}
\end{figure}

\begin{figure}[h!]
\begin{center}
\includegraphics[width=0.4\textwidth]{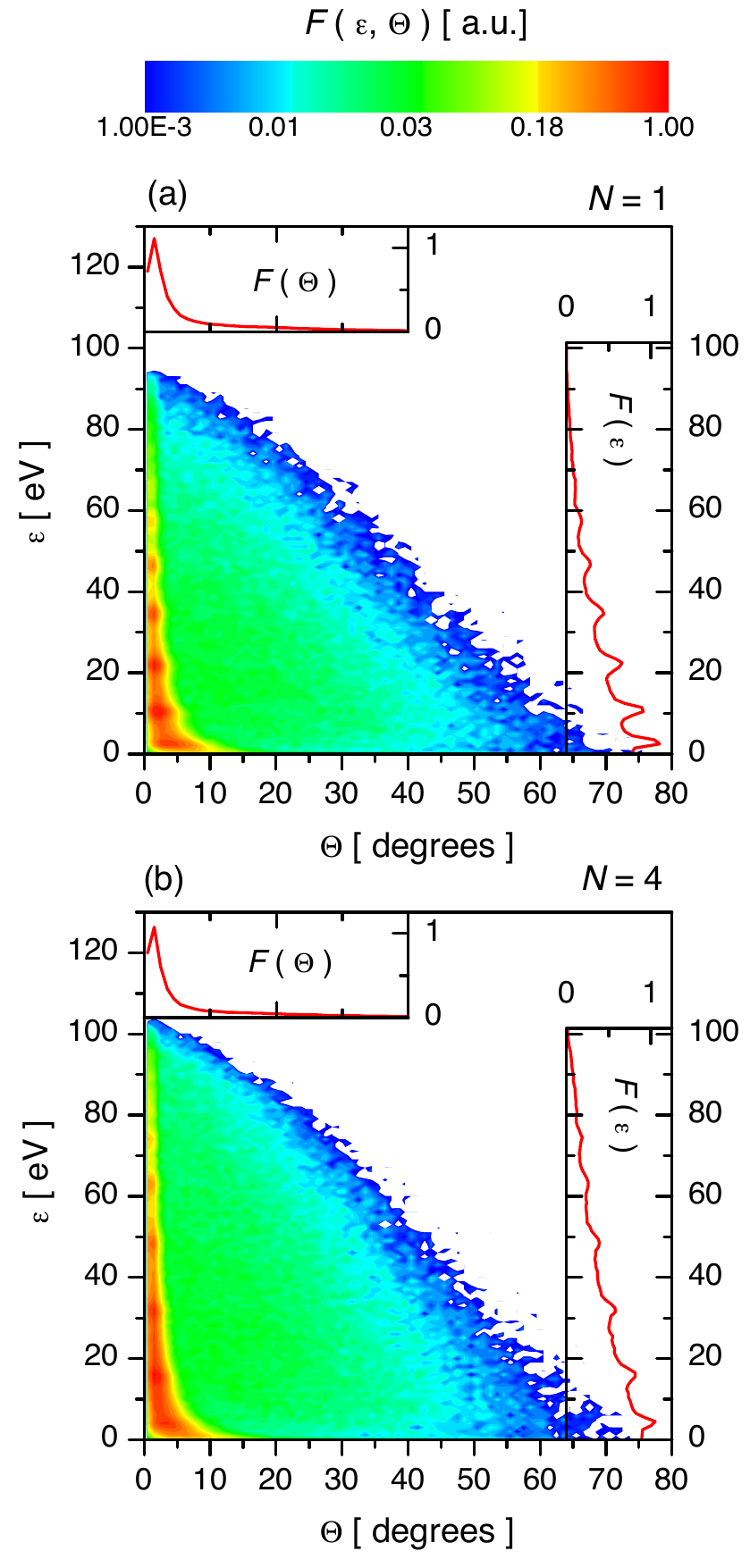}
\caption{Energy and angular distribution of the O$_2^+$ ion flux, $F(\varepsilon,\Theta)$ (in arbitrary units) at the grounded electrode in discharges driven by sawtooth-down waveforms with $N$ = 1 (a) and $N$ = 4 (b) harmonics. The insets illustrate the energy and angular distributions (respective integrals of $F(\varepsilon,\Theta)$ according to incidence angle and energy, respectively.) All distributions have been normalised to a maximum value of 1, for easier comparison. $\phi_{\rm pp}$ = 400 V and $p$ = 20 Pa.}
\label{fig:sawtooth_angular}
\end{center}
\end{figure}

Figure~\ref{fig:sawtooth_energy_flux} shows the mean ion energy at different pressures, for different number of harmonics. A comparison of the results at different pressures confirms the expected behaviour that the mean ion energy decreases with increasing pressure (due to the varying collisionality of the sheaths). The data, on the other hand, indicate a marginal effect of $N$ on the mean energy of the ions, at any fixed pressure. As explained in section \ref{sec:valleys} a "reversed" waveform (in the present case sawtooth down/up) mirrors the plasma on the symmetry mid-plane, this way the difference between the mean energies at the powered and grounded electrodes represents the control range of $\langle \varepsilon \rangle$ by phase control. Such a control, i.e. achieving different ion energies at the two electrodes, as inferred from figure~\ref{fig:sawtooth_energy_flux}, is not feasible with sawtooth-type waveforms. Furthermore, as figure \ref{fig:sawtooth_angular0}(a) reveals, the angular distribution function (IADF) of the ions is also rather insensitive to the value of $N$. At 2 Pa pressure a slight influence of $N$ is found at small angles, whereas at 5 Pa the IADFs are practically identical at various numbers of harmonics, except for $N$ = 1. This behaviour  originates from the fact that the maximum sheath widths ($s$) and the voltage drops over the sheaths are weakly influenced by the operating conditions. As confirmed in figure \ref{fig:sawtooth_angular0}(b), e.g., the maximum width of the sheath at the grounded electrode, $s_{\rm g}$, is very nearly the same for all conditions. The energy- and angle-resolved distributions of the ions at the grounded electrode, $F(\varepsilon,\Theta)$, show small differences when the number of harmonics is changed, as illustrated in figure \ref{fig:sawtooth_angular} for the case of 20 Pa pressure. The peaks in the IFEDF become less pronounced when the number of excitation harmonics is increased. This is caused by the reduction of the size of those domains in space and time, where ions that have undergone a charge exchange collision, can accumulate, similarly to the case of the dual-frequency waveform analysed in figure~\ref{fig:DF_EDDI}.

\section{Summary}

In this work, we have investigated the properties of low-pressure oxygen CCPs, with focus on the ion properties: possibilities of controlling the mean energy of ions bombarding the electrode surfaces, the flux-energy, the angular- as well as the joint energy-angular distribution of these ions. These characteristics have been studied for various excitation waveforms, including single-frequency, classical dual-frequency, valleys-type and sawtooth-type waveforms.  

In the case of single-harmonic excitation the shape of the energy-angular distribution at low pressures was understood by a simple analytical model. A more elaborated model was applied to account for the shape of the IFEDF at dual-frequency excitation. 

A remarkable difference has been found between the distributions emerging under classical DF and valleys-type excitation: in the case of the classical DF excitation the application of the low-frequency voltage component was found to increase the width of the electrode sheaths significantly, thereby increasing their collisionality. In the case of the valleys-type waveforms the enhancement of the mean ion energy was found to occur without this effect, and as a result a more confined beam of ions (in terms of angular spread) was found to reach the electrode surfaces. Hence, applying valley-type waveforms allows for a control of the ion energy without strongly affecting the narrow ion angular distribution function. In the case of sawtooth-type waveforms a weak effect of the waveform shape (defined by the number of harmonics, $N$) on the angular distributions was found. 

The electronegativity in the presence of different driving voltage waveforms was studied as a function of the voltage amplitudes and the number of harmonics. Its changes were understood based on the electron power absorption dynamics and changes of the mean electron energy that strongly affects the electronegativity via the energy dependence of the attachment cross section.

The complex waveforms, especially the classical dual-frequency waveform were found to "smoothen" the IFEDFs by making disappear the wide spatio-temporal domains within the sheath region, where ions that have undergone a charge exchange collision accumulate. The application of such waveforms, is therefore advantageous whenever a smooth IDEDF is preferred. When, on the other hand, a distinct peak in the IFEDF would be required, the method presented in \cite{Eddi15} may be followed. 

Although our discharge model and its numerical implementation have been benchmarked with various experimental data for a wide range of operating conditions and excitation waveforms \cite{Derzsi2016,Derzsi2017,DonkoEPS2018}, additional experimental data for multi-frequency discharges (e.g. on electron density, electronegativity, singlet delta molecule density, etc.) would be desired to allow further refining and verification of the discharge model. Such data, at present, exist mostly for single-frequency discharges \cite{Kullig}, data for dual-frequency discharges are limited \cite{newref1,newref2}, to our best knowledge.

\ack This work has been supported by National Research, Development and Innovation Office of Hungary (NKFIH, K119357, PD-121033), by the US National Science foundation (PHY 1601080), the DFG via SFB TR87, the International Joint Research Promotion Program (Type A) of Osaka University, the JSPS Grants-in-Aid for Scientific Research (S)15H05736 and the J. Bolyai Scholarship of the Hungarian Academy of Sciences (AD).

\end{document}